\documentclass{article}
\usepackage[latin1]{inputenc}
\usepackage[english]{babel}
\usepackage{dsfont}
\usepackage[cyr]{aeguill}
\usepackage{amsmath,amssymb}
\usepackage{amsthm}
\usepackage{eurosym}
\usepackage{multicol}
\usepackage{color}
\usepackage{fancyhdr}
\usepackage[pdftex]{graphicx}
\usepackage[pdftex]{hyperref}
\usepackage{verbatim}
\usepackage{setspace}
\usepackage{stmaryrd}

\newcommand{\proba}{\mathbb{P}} 
\newcommand{\indic}{\mathds{1}} 
\newcommand{\E}{\mathbb{E}} 
\newcommand{\Cov}{\text{Cov}} 
\newcommand{\Var}{\text{Var}} 
\newcommand{\corr}{\text{corr}} 

\theoremstyle{plain}
\newtheorem{thm}{Theorem}

\providecommand{\keywords}[1]{\textbf{\textit{Keywords --}} #1}

\title{Fractal properties, information theory, and market efficiency}

\author{Xavier Brouty$^{\text{a}}$, Matthieu Garcin$^{\text{b,}}$\thanks{Corresponding author: matthieu.garcin@m4x.org. \newline $^{\text{a}}$ ESILV, 92916 Paris La Défense, France. \newline $^{\text{b}}$ Léonard de Vinci Pôle Universitaire, Research center, 92916 Paris La Défense, France.}} 

\date{\today}

\begin{document}

\maketitle

\begin{abstract}
Considering that both the entropy-based market information and the Hurst exponent are useful tools for determining whether the efficient market hypothesis holds for a given asset, we study the link between the two approaches. We thus provide a theoretical expression for the market information when log-prices follow either a fractional Brownian motion or its stationary extension using the Lamperti transform. In the latter model, we show that a Hurst exponent close to $1/2$ can lead to a very high informativeness of the time series, because of the stationarity mechanism. In addition, we introduce a multiscale method to get a deeper interpretation of the entropy and of the market information, depending on the size of the information set. Applications to Bitcoin, CAC 40 index, Nikkei 225 index, and EUR/USD FX rate, using daily or intraday data, illustrate the methodological content.
\end{abstract}




\section{Introduction}

One of the main challenges for quantitative asset managers is to find statistical models able to predict future price evolutions of financial assets. In this perspective, the classical financial theory is discouraging. Indeed, the efficient market hypothesis (EMH) states that current prices reflect all the available information, so that statistical arbitrage is impossible. The repeated profits of many actors in this industry of course disprove the EMH, at least as a general statement. In practice, the difficulty of predicting future prices is not shared equally by all financial assets. It is thus important to have statistical methods able to identify, among several time series of prices, which can successfully be forecast and which cannot. A rich and varied literature, mainly in econophysics, therefore addresses this question. Two of the most widespread approaches are based on information theory and on fractals.

The main tool of information theory is Shannon's entropy. It measures the uncertainty inherent to a specific probability distribution. The distribution with the highest uncertainty, that is the one leading to the biggest difficulty of predicting the value of a random variable following this distribution, is the uniform distribution. The usefulness of the entropy notion in the analysis of the EMH thus appears when one is able to define probability distributions that are uniform if and only if the EMH holds. It is for instance the case of ordinal patterns of a time series of prices, which lead to the permutation entropy~\cite{BPompe}. A more natural method focuses on the conditional distribution of the sign of future price returns~\cite{Risso,BG}. We will study the latter entropy, from which we build the market information. This information consists of the difference between the measured entropy and an ideal entropy consistent with the EMH~\cite{BG,GarcinComplexity}.

The fractal analysis is an alternative approach used for determining the degree of market efficiency of a financial asset. Mainly based on the Hurst exponent, this method studies the impact of changing the time scale of the observations on some statistic of a time series. It asserts that the EMH corresponds to a Hurst exponent equal to $1/2$~\cite{Peters}. When one uses a moment-based estimator of the Hurst exponent, this approach implicitly relies on the assumption that log-prices follow a fractional Brownian motion (fBm). This model, which uses fractional calculus to reproduce specific fractal properties~\cite{MvN}, is consistent with the interpretation of the Hurst exponent regarding the EMH and given above, but it relies on strong assumptions, like Gaussian price returns.

It is tempting to compare the outcome of the two competing approaches cited above, namely entropy-based and fractal methods. Are they redundant, do they complement each other? With the aim of combining the two approaches, the multiscale entropy (MSE) makes it possible to study the entropy of a system observed at various time scales~\cite{CGP2002,CGP2005}. It has been used in particular to analyse the scale-dependent market efficiency~\cite{MRE,ARRA}. Beyond this attempt to build a synthetic method, some rare articles examine the theoretical link between entropy and fractal dimension~\cite{ZDV}, or the link, obtained by simulations, between the MSE and the Hurst exponent of an fBm~\cite{MDK}. It is also possible to obtain a theoretical expression for the permutation entropy in the case where the data are generated by some specific stochastic process, including the fBm~\cite{BS,ZPM}.

In this article, we intend to study the link between information theory and fractal properties, in the perspective of investigating the EMH, in three different and new directions. First, in addition to the MSE, we study theoretically and empirically the dependence of both the entropy and the market information on the size of the information set. Second, we explore the theoretical link between the Hurst exponent of an fBm and its market information, using the definition of entropy based on the conditional distribution of the sign of future price returns, instead of the sole permutation entropy. Third, noting that the fBm is not the only fractional model having fractal properties, we derive similar results for a stationary extension of the fBm, namely its inverse Lamperti transform. This kind of result is useful for instance for modelling fixed-income instruments and it shows new possible interpretations of the Hurst exponent regarding the information of time series.

Overall, this article sheds new lights on the informational content of fractional models. It confirms in a new manner previous results on the forecast of fBm in finance~\cite{GNR,GMR,GarcinForecast}. It also deepens the understanding of the stationary inverse Lamperti transform of the fBm~\cite{GarcinLamperti}. In this stationary framework, we get a counter-intuitive result, because a Hurst exponent equal to $1/2$ can depict a very high information in the series, depending on the intensity of the mean reversion. All these results may be instructive for quantitative traders, risk managers, and every person interested in forecasting time series in finance or in another context.

The rest of the paper is organized as follows. Section~\ref{sec:marketInfo} presents the entropy-based concept of market information. In Section~\ref{sec:infofract}, we derive theoretical results about the market information of two fractional processes: the fBm and its stationary extension with the Lamperti transform. We present in Section~\ref{sec:MSinfo} the multiscale approach applied to market information along with numerical results for both simulated time series and real financial data, including stock indices, Bitcoin, and an FX rate. Section~\ref{sec:conclu} concludes.

\section{Market information}\label{sec:marketInfo}

We are given a discrete random vector $(X_1,...,X_L)'$. We assume that each component of this vector has its value in $\{0,1\}$, so that the vector can be in one of the $2^L$ possible states noted $\textbf{s}^L_i$, for $i\in\llbracket 1,2^L\rrbracket$. In information theory, Shannon's entropy is a basic concept which quantifies the amount of uncertainty in a probability distribution. This uncertainty can also be seen as an aspect of the complexity of this distribution~\cite{LMC,GarcinComplexity}. The Shannon entropy of $(X_1,...,X_L)'$ is defined by~\cite{CT}:
$$H(X_1,...,X_L)=-\sum_{i=1}^{2^L}\proba((X_1,...,X_L)'=\textbf{s}^L_i)\log_2\left(\proba((X_1,...,X_L)'=\textbf{s}^L_i)\right).$$
The least informative probability distribution, that is the one with the highest uncertainty, is the uniform distribution. In this case, the entropy is $L$. The lowest uncertainty corresponds to the degenerate distribution in which all the probability is concentrated in a single state. Using the convention that $0\log_2(0)=0$, which we easily get from the limit of $x\log_2(x)$ when $x\rightarrow 0$, this degenerate case leads to a zero entropy. All the probability distributions have an entropy in the interval $[0,L]$, thus delimited by these two special cases.

Shannon's entropy has been exploited in finance in several ways, in particular to determine whether the market is efficient or not. We can identify two main lines of research, depending on how one defines the random vector $(X_1,...,X_L)'$ and its corresponding probability. First, the Bandt and Pompe method focuses on the distribution of the ordinal patterns of a time series of prices, giving rise to what is known as the permutation entropy~\cite{BPompe,ZTS,ZBB,BFF}. The second approach, which is the one we are interested in, is the Risso method. It applies the entropy to the discrete distribution of the sign of successive price returns~\cite{Risso,MBM,Ducournau,BG}. More precisely, we consider a sequence of prices of a given financial asset, $P_0,...,P_n$, sampled at a daily frequency.\footnote{ The daily frequency is chosen to fix the ideas, but the approach can naturally be extended to other frequencies.} From these prices, we compute price returns over $m$ days, for each date $i\in\llbracket m,n\rrbracket$,
$$R_{m,i}=\frac{P_{i}-P_{i-m}}{P_{i-m}},$$
and we build a new sequence of random variables, $J_{m,i}$, which are indicators of positive price return:
$$J_{m,i}=\left\{\begin{array}{ll}
1 & \text{if } R_{m,i}> 0 \\
0 & \text{else.}
\end{array}\right.$$
Assuming that the random vector $\mathcal J_i=(J_{m,i},J_{m,i+m},J_{m,i+2m},...,J_{m,i+Lm})'$, of size $L+1$, is part of a stationary sequence $\{\mathcal J_i\}_{i\in\llbracket m,n-Lm\rrbracket}$, we note the entropy of this vector
\begin{equation}\label{eq:entropymarket}
H^{L+1}_m=H(J_{m,.},...,J_{m,.+Lm}).
\end{equation}
If we know the distribution of a future increase indicator $J_{m,.+Lm}$, conditionally on the previous $L$ indicators $J_{m,.},...,J_{m,.+(L-1)m}$, we can also write this entropy
\begin{equation}\label{eq:entropymarketbis}
H^{L+1}_m=-\sum_{i=1}^{2^L}\sum_{j=1}^{2}\pi_m^L(\textbf{s}^1_j|\textbf{s}^L_i)p_m^L(\textbf{s}^L_i)\log_2\left(\pi_m^L(\textbf{s}^1_j|\textbf{s}^L_i)p_m^L(\textbf{s}^L_i)\right),
\end{equation}
where $p_m^L(\textbf{s}^L_i)=\proba((J_{m,.},...,J_{m,.+(L-1)m})'=\textbf{s}^L_i)$ and $\pi_m^L(\textbf{s}^1_j|\textbf{s}^L_i)=\proba(J_{m,.+Lm}=\textbf{s}^1_j|(J_{m,.},...,J_{m,.+(L-1)m})'=\textbf{s}^L_i)$. Using the notation of the conditional entropy
\begin{equation}\label{eq:conditionalEntropy}
H(J_{m,.+Lm}|J_{m,.},...,J_{m,.+(L-1)m})=-\sum_{i=1}^{2^L}p_m^L(\textbf{s}^L_i)\sum_{j=1}^{2}\pi_m^L(\textbf{s}^1_j|\textbf{s}^L_i)\log_2\left(\pi_m^L(\textbf{s}^1_j|\textbf{s}^L_i)\right),
\end{equation}
we can decompose equation~\eqref{eq:entropymarketbis} thanks to the chain rule~\cite[Th.2.2.1]{CT}:
$$H(J_{m,.},...,J_{m,.+Lm})=H(J_{m,.},...,J_{m,.+(L-1)m})+H(J_{m,.+Lm}|J_{m,.},...,J_{m,.+(L-1)m}),$$
that is
\begin{equation}\label{eq:chainrule}
H^{L+1}_m=H^L_m+H(J_{m,.+Lm}|J_{m,.},...,J_{m,.+(L-1)m}).
\end{equation}
In the case where the market is efficient, we have $\pi_m^L(1|\textbf{s}^L_i)=\proba(J_{m,.+Lm}=1)=1/2$: the uncertainty of a future price increment is maximal, whether or not we condition on past observations. We note $J^{\star}_{m,.+Lm}$ this theoretical EMH-based future price increment indicator, which is independent from the vector $(J_{m,.},...,J_{m,.+(L-1)m})'$. The corresponding entropy is
\begin{equation}\label{eq:chainrulebis}
\begin{array}{ccl}
H^{L+1,\star}_m & = & H(J_{m,.},...,J_{m,.+(L-1)m},J^{\star}_{m,.+Lm}) \\
 & = & H(J_{m,.},...,J_{m,.+(L-1)m})+H(J^{\star}_{m,.+Lm}|J_{m,.},...,J_{m,.+(L-1)m}) \\
 & = & H(J_{m,.},...,J_{m,.+(L-1)m})+H(J^{\star}_{m,.+Lm}) \\
 & = & H^L_m+1.
\end{array}
\end{equation}
where we successively used the chain rule of the conditional entropy, the independence of $J^{\star}_{m,.+Lm}$ with respect to the vector $(J_{m,.},...,J_{m,.+(L-1)m})'$, and the 2-state uniformity of $J^{\star}_{m,.+Lm}$. 

Finally, the market information is the divergence between the uncertainty in the real financial time series and the uncertainty in the ideal efficient time series~\cite{BG}:
\begin{equation}\label{eq:marketInfo}
\begin{array}{ccl}
I^{L+1}_m & = & H^{L+1,\star}_m-H^{L+1}_m \\
 & = & 1-H(J_{m,.+Lm}|J_{m,.},...,J_{m,.+(L-1)m}),
\end{array}
\end{equation}
according to equations~\eqref{eq:chainrule} and~\eqref{eq:chainrulebis}. The market information is always positive, being equal to zero if and only if $\forall i\in\llbracket 1,2^L\rrbracket$, $\pi_m^L(1|\textbf{s}^L_i)=1/2$. One can use this quantity to build statistical tests of market efficiency~\cite{BG,SMM}. Estimating this quantity is thus a major challenge. If we simply replace the conditional probabilities by their empirical estimator in equations~\eqref{eq:marketInfo} and~\eqref{eq:conditionalEntropy}, we get a plug-in estimator which converges toward the true market information when $n\rightarrow\infty$~\cite{Verdu}. However, this estimator has two drawbacks. First, it is non-asymptotically biased. The bias is particularly striking when the market is really efficient, that is with a theoretical information equal to zero, because the estimated information is always positive. Second, when $L$ is large, the estimation of the empirical probabilities requires a huge amount of data, much larger than the number of states $2^{L+1}$. More advanced estimators make it possible to bypass these two drawbacks~\cite{GKB,Verdu}. One can for example cite estimators based on the Lempel-Ziv compression algorithm~\cite{KAS}, on the context-tree weighting algorithm~\cite{GKB}, or on Bayesian context trees~\cite{PaKo}.

We also note that, from the beginning, we have considered variables with only two possible states, 0 and 1. The definitions above can be generalized to a higher number of states~\cite{SLOS,LBA}, but the interpretation regarding the market efficiency is then less obvious~\cite{BG,GarcinComplexity}.

\section{Market information of fractal dynamics}\label{sec:infofract}

Fractal dynamics are often characterized by a selfsimilarity property. A stochastic process $X_t$ is said to be $\mathcal H$-selfsimilar if, whatever $c>0$ and $t\geq 0$, $X_{t}\overset{d}{=}c^{-\mathcal H}X_{ct}$, where $\overset{d}{=}$ means equality in finite-dimensional distributions. The parameter $\mathcal H$ is the Hurst exponent. The Brownian motion is an example of a stochastic process having such a property, with $\mathcal H=1/2$. This process is widespread in finance, but it is only a particular case of the fBm, which offers more flexibility in the selection of the Hurst exponent. The fBm of Hurst exponent $\mathcal H$ is a $\mathcal H$-selfsimilar stochastic process which has some additional simple properties: a Gaussian distribution, stationary increments. Some extensions and transformations of the fBm make it possible to consider more general dynamics with some fractal features or even with more complex multifractal properties~\cite{SSL,KY,DiMatteo,Grech}. One can cite for instance stationary transformations~\cite{CKM,HN,GarcinLamperti,GarcinEstimLamp}, non-Gaussian extensions~\cite{ST,WBMW,AG}, Hurst exponents varying through time in a deterministic~\cite{Coeurjolly,BP10,Garcin2017} or stochastic fashion~\cite{BPP,BP,GarcinMPRE}, and the multifractal random walk~\cite{BDM,BDMfin,MMA}.

In what follows, we are interested in the entropy analysis for dynamics of log-prices generated by an fBm or by the stationary delampertized fBm. We note that the results below, in particular Theorems~\ref{th:infofBm} and~\ref{th:infoLamperti}, are also valid if the price, instead of being the exponential of a specific fractal process, is any monotonic function of this process, as soon as this function remains the same trough time. After giving some details on the models, we provide a theoretical formula for the market information $I^2_m$, that is the ability of a single lagged observation to predict future price evolutions.

\subsection{Log-prices following an fBm}\label{sec:fBm}

Exploiting the fractional derivative or the fractional integral of a standard Brownian motion, Mandelbrot and Van Ness have proposed the fBm, a simple Gaussian model with stationary increments which is able to fit the fractal property of log-prices~\cite{MvN}. More precisely, an fBm of Hurst exponent $\mathcal H\in(0,1)$ and volatility parameter $\sigma>0$ is a stochastic Gaussian process $B^{\mathcal H}_t$ such that $B^{\mathcal H}_0=0$, $\E\{B^{\mathcal H}_t\}=0$, and 
\begin{equation}\label{eq:defFBM}
\E\left\{\left(B^{\mathcal H}_t-B^{\mathcal H}_s\right)^2\right\}=\sigma^2|t-s|^{2\mathcal H}
\end{equation}
for all $s,t\in\mathbb R$. It is worth noting that the case $\mathcal H=1/2$ corresponds to the standard Brownian motion. On the contrary, when $\mathcal H<1/2$ (respectively $\mathcal H>1/2$), $B^{\mathcal H}_t$ is the fractional derivative (resp. integral) of order $1/2-\mathcal H$ (resp. $\mathcal H-1/2$) of a standard Brownian motion and is therefore not Markovian. In other words, as soon as $\mathcal H\neq 1/2$, it is legitimate to forecast future evolutions of an fBm~\cite{NP}. In practice, one can do it by exploiting the covariance of the process, which we easily get from equation~\eqref{eq:defFBM}:
\begin{equation}\label{eq:covFBM}
\E\left\{B^{\mathcal H}_tB^{\mathcal H}_s\right\}=\frac{\sigma^2}{2}\left(|t|^{2\mathcal H}+|s|^{2\mathcal H}-|t-s|^{2\mathcal H}\right),
\end{equation}
for all $s,t\in\mathbb R$. In finance, the propensity of the fBm to be forecast contradicts the EMH and is at the root of many statistical arbitrage strategies~\cite{GNR,GMR,GarcinForecast}. It is thus relevant to quantify the market information of such a dynamic. Using the entropy-based method exposed in Section~\ref{sec:marketInfo} we specify the theoretical link between the market information and the Hurst exponent in the following theorem.

\begin{thm}\label{th:infofBm}
Let $B^{\mathcal H}_t$ be an fBm of Hurst exponent $\mathcal H$ and volatility parameter $\sigma$. Let the price $P_t$ be equal to $\log(B^{\mathcal H}_t)$ for all $t\in\mathbb R$, and $m>0$. Then, the market information $I_m^2$ introduced in equation~\eqref{eq:marketInfo} is equal to
\begin{equation}\label{eq:infofBm}
I_m^2=1+f\left(\frac{1}{2}-\frac{1}{\pi}\arctan\left(\frac{\rho}{\sqrt{1-\rho^2}}\right)\right)+f\left(\frac{1}{2}+\frac{1}{\pi}\arctan\left(\frac{\rho}{\sqrt{1-\rho^2}}\right)\right),
\end{equation}
where $f:x>0\mapsto x\log_2(x)$ and $\rho=2^{2\mathcal H-1}-1$.
\end{thm} 

The proof of Theorem~\ref{th:infofBm} is postponed in Appendix~\ref{sec:proof_infofBm}. We note that the information and the entropy in this theorem are based on probabilities which are the same as the ones appearing in the permutation entropy~\cite{BS}.

\begin{figure}[htbp]
	\centering
		\includegraphics[width=0.6\textwidth]{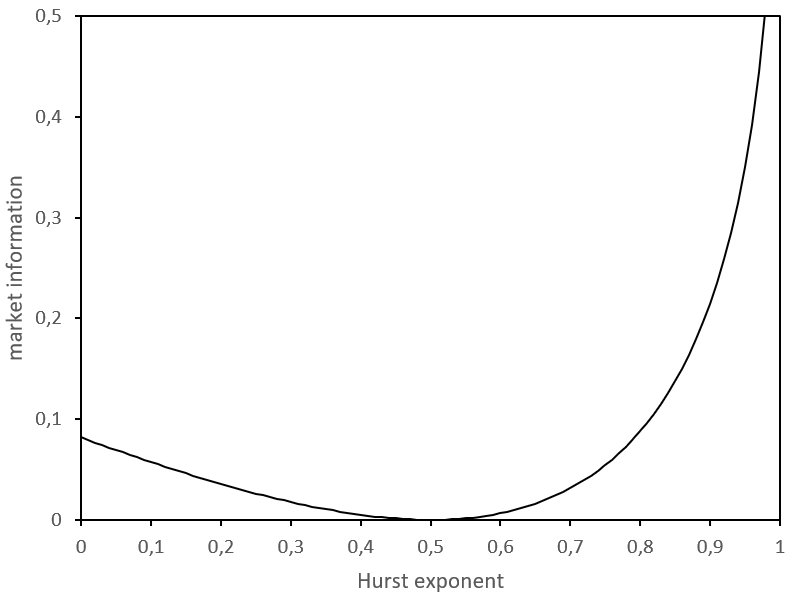} 
\begin{minipage}{0.7\textwidth}\caption{Theoretical market information $I^2_m$ for a dynamic of log-prices following an fBm, depending on its Hurst exponent, according to Theorem~\ref{th:infofBm}.}
	\label{fig:HurstInfo}
\end{minipage}
\end{figure}

Figure~\ref{fig:HurstInfo} represents the theoretical market information $I^2_m$ with respect to the Hurst exponent $\mathcal H$, according to Theorem~\ref{th:infofBm}. Consistently with the above remark, the case where $\mathcal H=1/2$, that is where the fBm is a standard (Markovian) Brownian motion, is the only case for which the market information is equal to zero. When $\mathcal H$ moves away from $1/2$, from above or from below, the market information increases. The increase of the information is stronger for $\mathcal H>1/2$ than for $\mathcal H<1/2$. In other words, having $\mathcal H\neq 1/2$ is enough for predicting future price increments, but the accuracy of the forecast is lower for $\mathcal H<1/2$ than for $1-\mathcal H$. It confirms previous findings based on another accuracy metric, namely the mean squared error or the hit ratio~\cite{NP,GarcinForecast}.

Theorem~\ref{th:infofBm} also shows that the market information does not depend on the size of the time steps $m$ if the dynamic is generated by an fBm. If this model appropriately describes the market, predictions with daily price returns are as accurate as predictions with intraday observations, as soon as the number of lagged returns is the same. This is of course an idealized and somehow unrealistic view and practitioners do not use in general the same model to describe prices with a high- or a low-frequency sampling of time. Indeed, despite its interest compared to simpler models, the fBm is only another model and fails to describe everything. For this reason, people may use and stack alternative models featuring also some fractal component, as detailed in the preamble of Section~\ref{sec:infofract}. As a consequence, one can reasonably think that the link between the selfsimilarity parameter $\mathcal H$ and the market information is not always as obvious as depicted by Figure~\ref{fig:HurstInfo}, which is restricted to a specific model, the fBm.

\subsection{Log-prices following a stationary delampertized fBm}

Among the models with a fractal feature, the stationary adaptations of the fBm are of particular interest in finance. They can indeed be useful for describing time series reknown to be stationary, like interest rates~\cite{ABKZ,MN}, FX rates~\cite{GarcinLamperti,GarcinMPRE}, and volatilities~\cite{CV,BLP,GJR,GG}. Several specifications of such a stationary model exist, for example depending on the presence or not of long memory~\cite{CKM,GarcinLamperti}. We focus on the delampertized fBm, which has been shown to be a relevant empirical choice~\cite{GarcinLamperti} and which leads to tractable equations.

The delampertized fBm relies on the Lamperti transform, which makes it possible to transform a stationary process $Y_t$ in a $\mathcal H$-selfsimilar process, $(\mathcal L_{\mathcal H} Y)_t=t^{\mathcal H}Y_{\ln(t)}$, and reciprocally a $\mathcal H$-selfsimilar process $X_t$ in a stationary process, $(\mathcal L^{-1}_{\mathcal H} X)_t=e^{-\mathcal Ht}X_{\exp(t)}$~\cite{FBA}. This inverse Lamperti transform has a stationarising effect like the mean-reverting Ornstein-Uhlenbeck mechanism~\cite{GarcinLamperti}. We place ourselves in the particular case where the underlying selfsimilar process is an fBm $B^{\mathcal H}_t$ and we add a parameter $\theta>0$ to adjust the strength of the mean reversion. We thus introduce the stationary process $\mathcal X^{\mathcal H,\theta}$, defined by
\begin{equation}\label{eq:LampertiTransfo}
\mathcal X^{\mathcal H,\theta}_t=\left(\mathcal L^{-1}_{\mathcal H} B^{\mathcal H}\right)_{\theta t}.
\end{equation}

It has been shown that, in general, the delampertized fBm is not equal to the fractional Ornstein-Uhlenbeck process~\cite{CKM}. However, in the case where $\mathcal H=1/2$, the two processes are equal: a delampertized Brownian motion $X^{\frac{1}{2},\theta}_t$ is a (non-fractional) Ornstein-Uhlenbeck process and thus is defined by the following equation, which explicitly incorporates the mean reversion mechanism around 0:
$$d\mathcal X^{\frac{1}{2},\theta}_t=-\frac{\theta}{2}\mathcal X^{\frac{1}{2},\theta}_t dt + \sqrt{\theta}dB^{1/2}_t,$$
where $B^{1/2}_t$ is a Brownian motion with volatility parameter $\sigma$.

Like in the case of the fBm, Theorem~\ref{th:infoLamperti} provides the market information when the log-price follows a delampertized fBm. As mentioned in the preamble of Section~\ref{sec:infofract}, this theorem remains valid if, instead of writing the price $P_t=\log(\mathcal X^{\mathcal H,\theta}_t)$, we write $P_t=g(\mathcal X^{\mathcal H,\theta}_t)$ for any monotonic function $g$. This would legitimate describing an interest rate or a volatility, and not only a log-price, by the delampertized fBm $\mathcal X^{\mathcal H,\theta}_t$.

\begin{thm}\label{th:infoLamperti}
Let $B^{\mathcal H}_t$ be an fBm of Hurst exponent $\mathcal H$ and volatility parameter $\sigma$. Let $\mathcal X^{\mathcal H,\theta}_t$ be the delampertized fBm defined by equation~\eqref{eq:LampertiTransfo}, with parameter $\theta>0$. Let the price $P_t$ be equal to $\log(\mathcal X^{\mathcal H,\theta}_t)$ for all $t\in\mathbb R$, and $m>0$. Then, the market information $I_m^2$ introduced in equation~\eqref{eq:marketInfo} is equal to
$$I_m^2=1+f\left(\frac{1}{2}-\frac{1}{\pi}\arctan\left(\frac{\rho}{\sqrt{1-\rho^2}}\right)\right)+f\left(\frac{1}{2}+\frac{1}{\pi}\arctan\left(\frac{\rho}{\sqrt{1-\rho^2}}\right)\right),$$
where $f:x>0\mapsto x\log_2(x)$, $\rho=\frac{2-h(2m\theta)}{4-2h(m\theta)}-1$, and $h:x\geq 0\mapsto 2\cosh(\mathcal Hx)-\left(2\sinh\left(\frac{x}{2}\right)\right)^{2\mathcal H}$.
\end{thm} 

The proof of Theorem~\ref{th:infoLamperti} is postponed in Appendix~\ref{sec:proof_infoLamperti}.

\begin{figure}[htbp]
	\centering
		\includegraphics[width=0.45\textwidth]{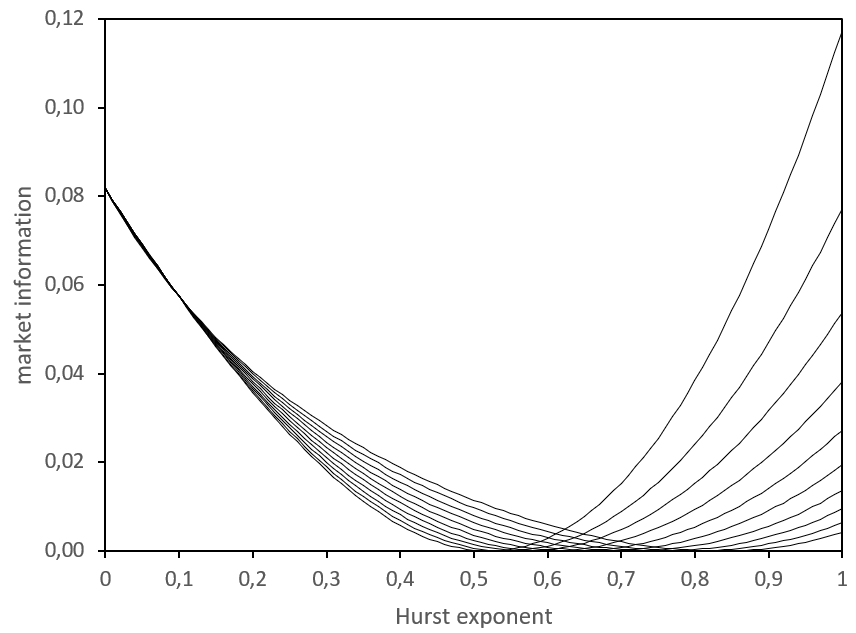} 
		\includegraphics[width=0.45\textwidth]{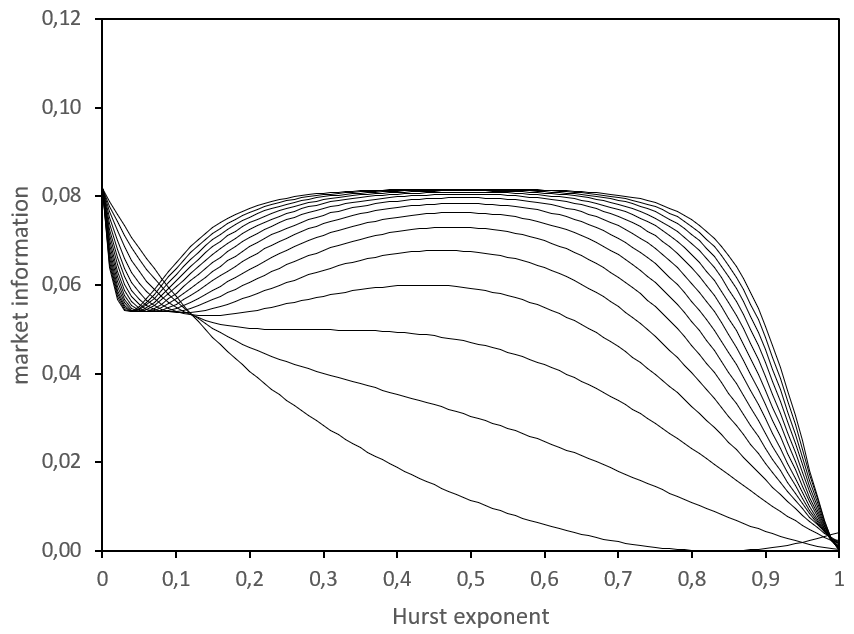} 
\begin{minipage}{0.7\textwidth}\caption{Theoretical market information $I^2_1$ for a dynamic of log-prices following a delampertized fBm, $\mathcal X^{\mathcal H,\theta}$, depending on its Hurst exponent. Each curve corresponds to a different value of $\theta$: between 0.1 and 1 by step of 0.1 (left graph, from bottom to top at the abscissa $\mathcal H=0.5$) and between 1 and 15 by step of 1 (right graph, from bottom to top at the abscissa $\mathcal H=0.5$).}
	\label{fig:HurstLampertiInfo}
\end{minipage}
\end{figure}

We display in Figure~\ref{fig:HurstLampertiInfo} the theoretical market information in the case of a delampertized fBm, following Theorem~\ref{th:infoLamperti}, for various values of the Hurst exponent $\mathcal H$ and of the mean reversion strength $\theta$. We obtain curves which have a shape which is in general very different from the fBm case, displayed in Figure~\ref{fig:HurstInfo}. In fact, the fBm case seems to be the limit of the delampertized fBm when $\theta$ tends to 0, that is when the fractional aspect is predominant compared to the mean reversion, or when $m$ tends to zero, that is when one considers the market information at a very high-frequency sampling. In fact, in Theorem~\ref{th:infoLamperti}, the combined parameter $m\theta$ is more relevant than  $\theta$ or $m$ alone and these two parameters play exactly the same role. For this reason, we have fixed $m=1$ in Figure~\ref{fig:HurstLampertiInfo}.

Like for the fBm, we observe that the market information strongly depends on the Hurst exponent, but, unlike the fBm case, the sampling $m$, or equivalently $\theta$, also plays an important role. We can identify two regimes, corresponding to the left and the right graphs in Figure~\ref{fig:HurstLampertiInfo}. For $\theta\in(0,1]$, increasing $\theta$ does not change much the market information for low $\mathcal H$, but it shrinks the information for high values of $\mathcal H$ and it translates the $\mathcal H$ minimizing the information, from $1/2$ to a higher value. On the other hand, when $\theta>1$, the minimum information is reached for $\mathcal H\rightarrow 1$ and, as soon as $\theta$ is large enough, the maximum information corresponds both to $\mathcal H\rightarrow 0$ and $\mathcal H=1/2$. 

This apparent contradiction with the traditional interpretation of the Hurst exponent is in fact clearly established in the literature devoted to this model~\cite{GarcinLamperti,GarcinEstimLamp}. The parameter $\mathcal H$ is related to the fractal properties of the process $B^{\mathcal H}_t$, not to the fractal properties of $\mathcal X^{\mathcal H,\theta}_t$. In other words, $\mathcal H$ is only the underlying Hurst exponent, not the Hurst exponent perceived when one conducts a selfsimilarity analysis on $\mathcal X^{\mathcal H,\theta}_t$. In fact, in the delampertized fBm $\mathcal X^{\mathcal H,\theta}_t$, two mechanisms are at work: the fractional aspect and the stationary aspect. The fractional aspect in the fBm, which generates its memory, leads to fractal properties, which are statistically observable at small time scales. The stationary aspect corresponds to a mean reversion statistically observable at larger time scales~\cite{GarcinLamperti,GG}. When one goes from a low time scale, that is a low $m$ or a low $\theta$, to a higher time scale, one gradually modifies the behaviour of $\mathcal X^{\mathcal H,\theta}_t$, from the pure fBm to the stationary process, and thus it modifies the way one predicts such a process, exploiting either the fractional aspect or the mean reversion.

This example of the delampertized fBm shows that the link between the market information and the Hurst exponent is not straightforward. These two quantities are in fact complementary notions.

\section{Multiscale market information}\label{sec:MSinfo}

In Section~\ref{sec:infofract}, we have studied the market information for time series with specific scaling properties. We are now interested in the scaling properties of the market information itself, for very general time series. We are going to present first the method of multiscale information analysis. An application to simulations and to real financial data will then follow.

\subsection{The method}\label{sec:methodmultiscale}

The multiscale aspect of the market information $I^{L}_m$ may lie either in its dependence on the number $L$ of lags considered, or in its dependence on the time scale $m$ of the price returns. These two axis provide different interpretations.

We first focus on the dependence of the market information on the parameter $L$. Before addressing this question regarding the market information, we have to understand the dependence of the entropy on $L$. This is the purpose of Theorem~\ref{th:concaventropy}.

\begin{thm}\label{th:concaventropy}
Let $m>0$ and $\mathcal J_i$ be the random vector $(J_{m,i},J_{m,i+m},J_{m,i+2m},...,J_{m,i+Lm})'$, where we assume that, whatever $L\in\mathbb N\setminus\{0\}$, the sequence $\{\mathcal J_i\}_{i\in\llbracket m,n-Lm\rrbracket}$ is stationary. Then, the mapping $L\mapsto H^L_m$, as defined in equation~\eqref{eq:entropymarket}, is increasing and concave in $L$.
\end{thm}

The proof of Theorem~\ref{th:concaventropy} is postponed in Appendix~\ref{sec:proof_concaventropy}.

The intuition behind Theorem~\ref{th:concaventropy} is that adding a lagged observation in the conditioning set diminishes the uncertainty about the future increase indicator of the price. But the more lagged observations one adds, the less informative each new observation. When, from a certain value of $L$, one adds uninformative lagged observations, the conditional entropy is constant from this $L$~\cite[Th.2.6.5]{CT} and $H^L_m$ becomes linear in $L$. 

In the many examples below, the entropy seems to be a linear function of $L$, but a more accurate look at  the first difference through the lens of the market information $I_m^L$ shows something different: $L\mapsto I_m^L$ is not a constant mapping for real financial datasets. This has important consequences. Indeed, since 
$$I^L_m=1+H^{L-1}_m-H^L_m,$$
the EMH holds whatever the size $L$ of the information set if and only if $\forall L\in\mathbb N\setminus\{0\}$, $I^L_m=0$, that is if and only if $L\mapsto H^L_m$ is linear of slope equal to 1. An inefficient market must thus have a nonlinear entropy $L\mapsto H^L_m$ or a slope lower than 1.

When the mapping $L\mapsto H^L_m$ is linear, its slope is similar to the entropy rate~\cite{CGP2005}, 
$$h=\underset{L\rightarrow\infty}{\lim} \frac{H^L_m}{L},$$
that is one minus the average amount of information carried by each vector of lagged observations. It is however difficult to estimate the entropy rate since it is based on a limit of the entropy when $L$ tends to infinity. Indeed, when the number of states, $2^L$, is big, that is not much lower than the number of observed returns $n-Lm$, there is a consequent statistical error in the estimation of the probabilities $p^L_m$ and $\pi_m^L$, which leads to underestimating the entropy by increasing artificially the divergence with respect to the uniform probability. This underestimation of the entropy is a challenge for estimating the entropy rate~\cite{CGP2005}, but it may also create spurious concavity of $L\mapsto H^L_m$ or lead to overestimate market information if one considers too large values for $L$.

If one restricts oneself to small values of $L$, one can also depict the marginal contribution to the information of each lagged observation thanks to the partial market information defined as the difference of informations of successive values of $L$: $\mathcal I^{L}_m=I^L_m-I^{L-1}_m$.

In the multiscale framework, the second dependence of interest for the market information is with respect to the parameter $m$. This is an adaptation to the information level of the MSE~\cite{CGP2002,CGP2005}. This concept of MSE has been used to study the complexity of financial time series~\cite{NW,LW,XS} and even to analyse the market efficiency at different time scales~\cite{MRE,ARRA}, noting that, consistently with the adaptive market hypothesis~\cite{Lo}, a market can be efficient at some time scales and not at others. Obviously, if the EMH is to be verified whatever $m$, then $\forall m>0$, $I^L_m=0$. But we don't have a unique pattern of the mapping $m\mapsto I^L_m$, when the EMH is violated. In particular, Section~\ref{sec:infofract} has shown that various models contradicting the EMH lead to a market information which may depend or not on the value of $m$.

\subsection{Simulations}\label{sec:Simul}

\begin{figure}[tbp]
	\centering
		\includegraphics[width=0.45\textwidth]{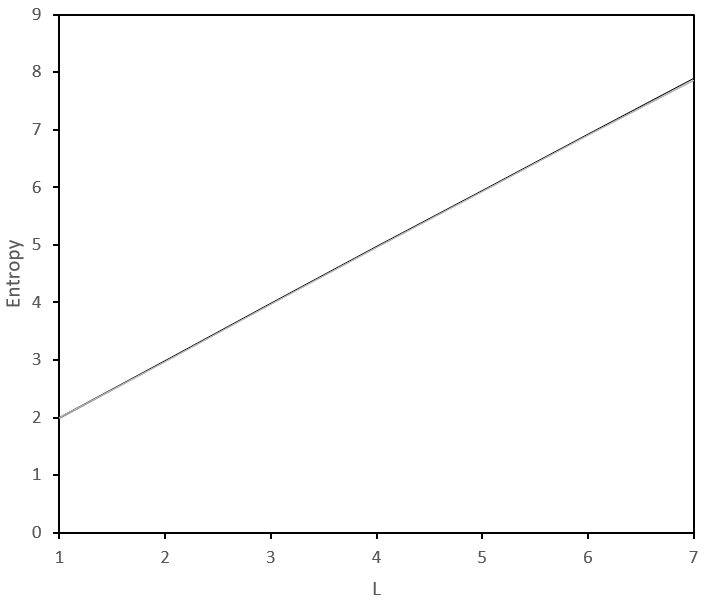} 
		\includegraphics[width=0.45\textwidth]{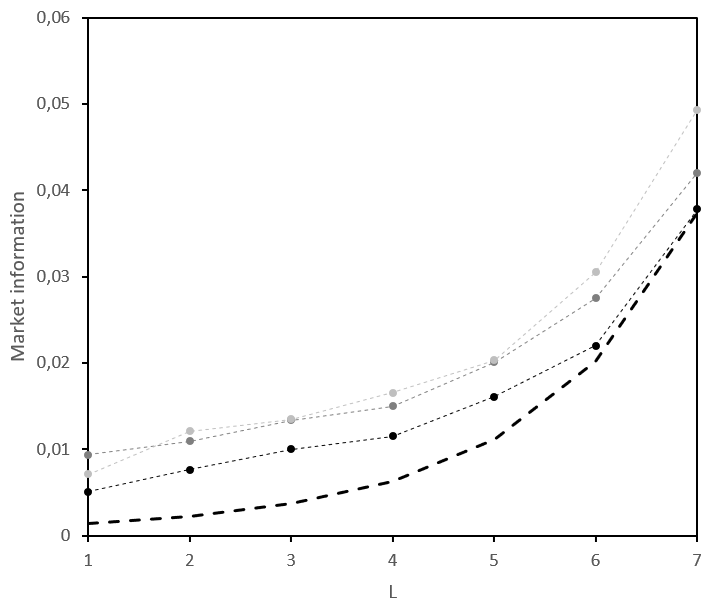} \\
		\includegraphics[width=0.45\textwidth]{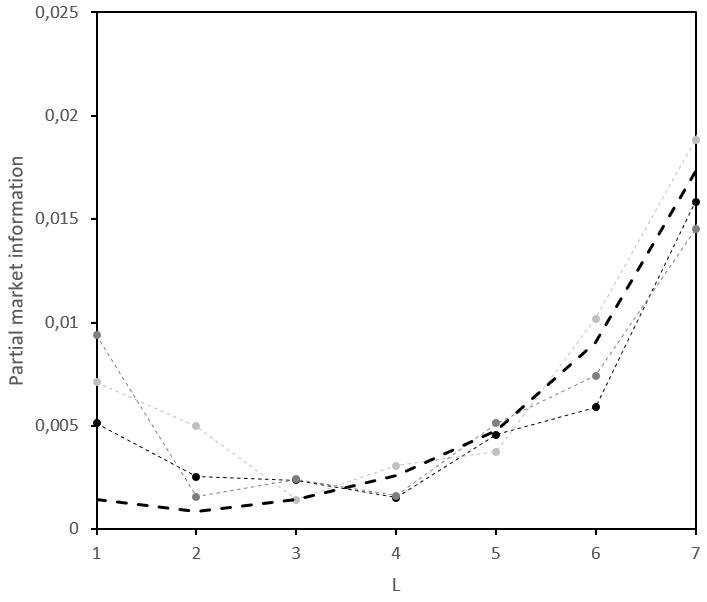}
		\includegraphics[width=0.45\textwidth]{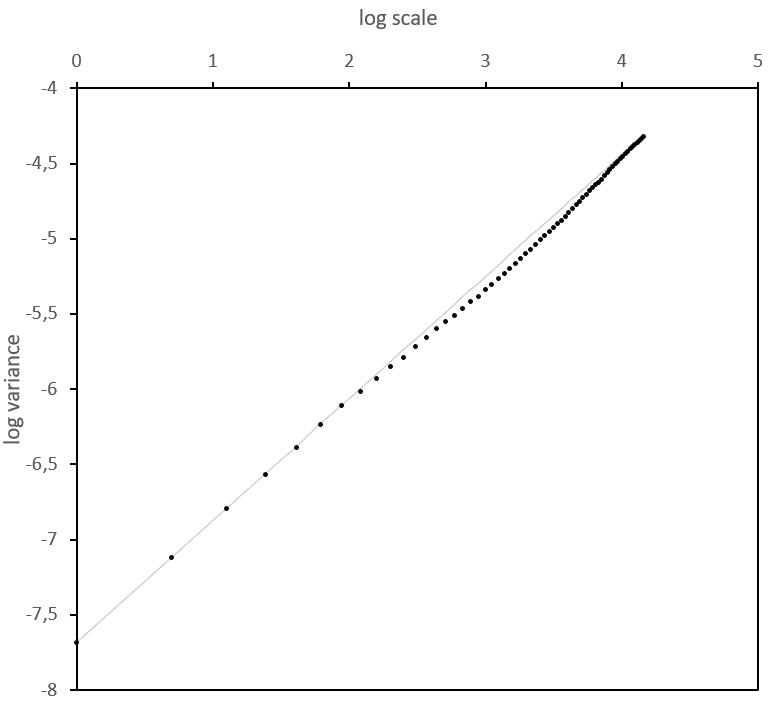}
\begin{minipage}{0.7\textwidth}\caption{Estimated entropy $H^{L+1}_m$ (top left), estimated market information $I^{L+1}_m$ (top right), and estimated partial market information $\mathcal I^{L+1}_m$ (bottom left) for $L\in\llbracket 1,7\rrbracket$ and $m$ equal to 1 (black), 2 (dark grey), and 3 (light grey). The thick dotted lines are derived from the $95\%$ confidence bound of a zero market information for $m=1$. Bottom right: Log-log plot (black) and linear model fitted on the smallest scales (grey). The price series is a simulated geometric fBm of Hurst exponent 0.4 and of length 3,000.}
	\label{fig:fBm}
\end{minipage}
\end{figure}

We simulate an fBm of Hurst exponent 0.4 and of length 3,000. We represent in Figure~\ref{fig:fBm} several indicators presented in this article. First, the estimated entropy $H^{L+1}_m$, as a function of the size $L$ of the vector considered, has a linear aspect, whatever the value of the time scale $m$. As exposed above, the linear aspect is not enough to assert the efficiency of a price following this dynamic. The slope is also to be considered. Its value is 0.98. The difference with the value consistent with the EMH, that is 1, is small and may be partly explained by the statistical error, but the same simulation with a Hurst exponent equal to 0.5 leads to a higher slope, equal to 0.99. More convincingly, the market information and the partial market information show a non-negligible contribution of lagged observations to the information of the series, whatever the time scale, in particular for more recent observations, that is small values of $L$. This is consistent with the propensity of such a dynamic to be forecast~\cite{GarcinForecast}. The larger partial information one observes for large values of $L$ illustrates the statistical artefacts described in Section~\ref{sec:methodmultiscale}. In order to disentangle the informative part from the statistical error, we display in Figure~\ref{fig:fBm} the $95\%$ confidence bound of a zero market information for $m=1$,\footnote{ Since increasing $m$ to 2 or 3 only diminishes the number of observations used in the estimation of the probabilities $p_m^L$ and $\pi_m^L$ by one or two units, the confidence bound for $m=2$ or 3 is very close to the case $m=1$.} using the asymptotic gamma distribution $\Gamma(k,\theta)$ of the empirical market information, which depends on the number of observations $n+1$ and on $L$, with $k=2^{L-1}$ and $\theta=[(n-mL)\ln(2)]^{-1}$~\cite{BG}. Moreover, by differentiating this bound, we get an idea of the lower bound for significant partial market information, as displayed with a thick dotted line in the corresponding graph, Figure~\ref{fig:fBm}. It shows that the three first lagged observations only seem to contribute significantly to the predictability of the sign of the future price return. Last, we also show the traditional log-log plot used to estimate the Hurst exponent, which is half the slope of the linear regression.

\begin{figure}[tbp]
	\centering
		\includegraphics[width=0.45\textwidth]{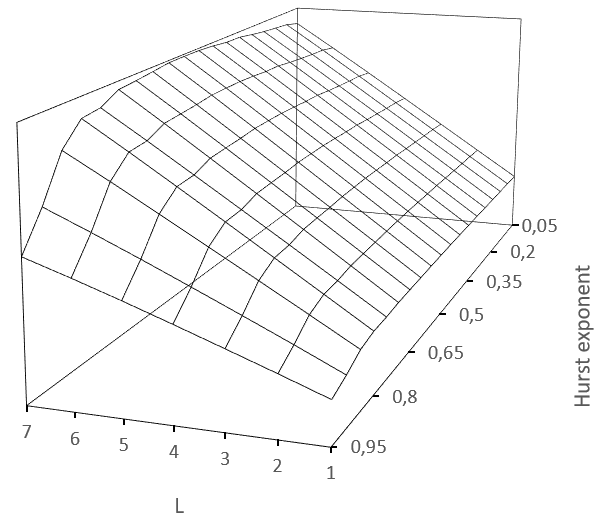} 
		\includegraphics[width=0.45\textwidth]{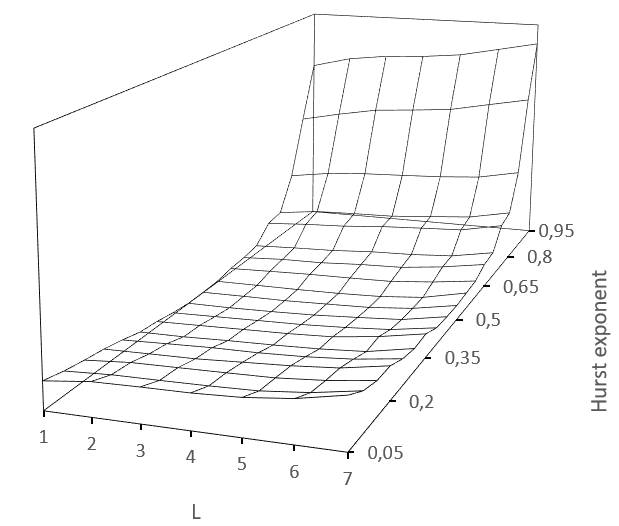} \\
		\includegraphics[width=0.45\textwidth]{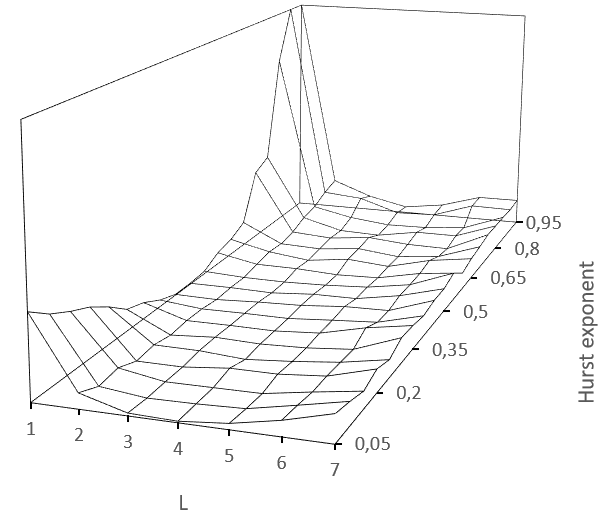}
		\includegraphics[width=0.45\textwidth]{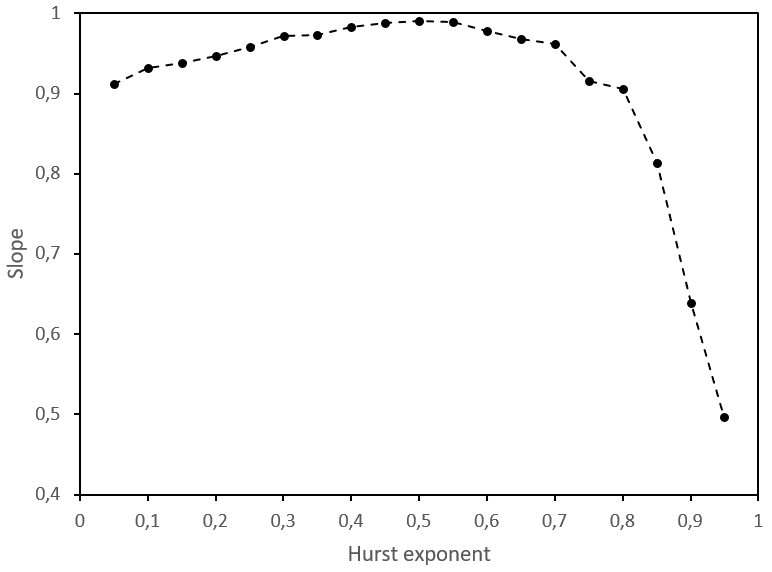}
\begin{minipage}{0.7\textwidth}\caption{Estimated entropy $H^{L+1}_m$ (top left), estimated market information $I^{L+1}_m$ (top right), and estimated partial market information $\mathcal I^{L+1}_m$ (bottom left) for $L\in\llbracket 1,7\rrbracket$, $m$ equal to 1, and for all the Hurst exponents in the interval $[0.05,0.95]$ with a step of $0.05$. Bottom right: Slope of the function $L\mapsto H^{L+1}$, depending on the Hurst exponent. Each price series is a simulated geometric fBm of length 3,000.}
	\label{fig:Surface}
\end{minipage}
\end{figure}

This simulation study of the fBm can be generalized to other values of $\mathcal H$. The results are gathered in Figure~\ref{fig:Surface} and confirm the theoretical analysis made in Section~\ref{sec:fBm}: the information is low when $\mathcal H$ is close to $1/2$, high when $\mathcal H$ is close to 0 and even higher when it is close to 1. We note however that, whatever $\mathcal H$, the mapping $L\mapsto H^L_m$ seems to be almost linear, with a slope all the more far from 1 as $\mathcal H$ is far from $1/2$. But it doesn't mean that the mapping is universally linear, whatever the model of dynamic.

\begin{figure}[tbp]
	\centering
		\includegraphics[width=0.45\textwidth]{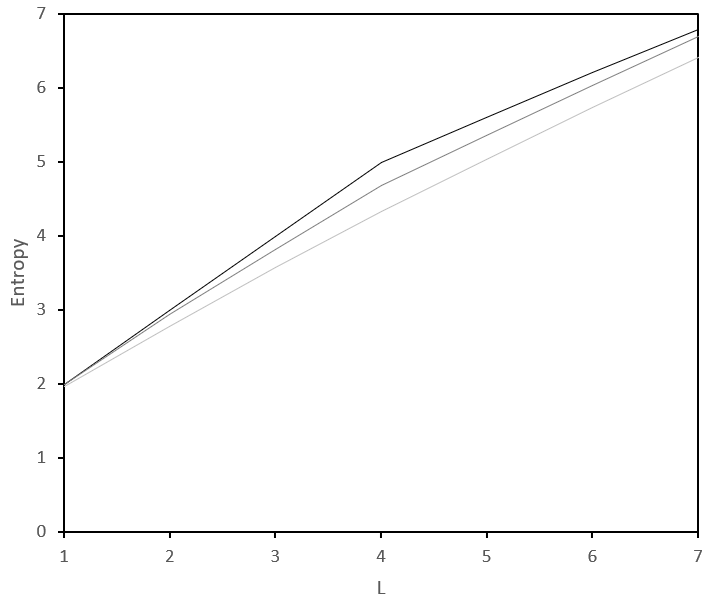} 
		\includegraphics[width=0.45\textwidth]{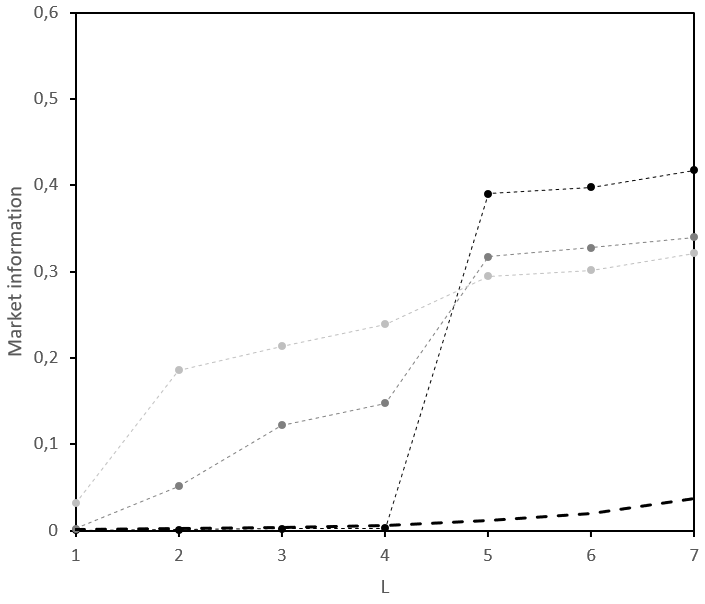} \\
		\includegraphics[width=0.45\textwidth]{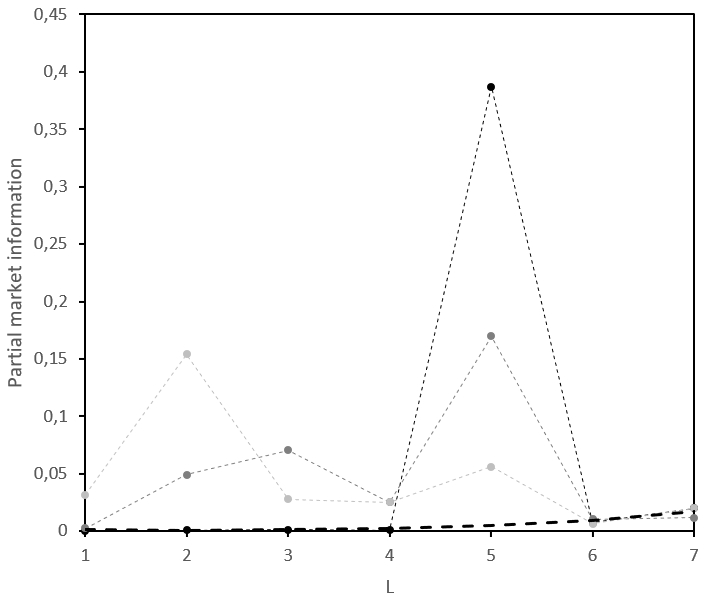}
		\includegraphics[width=0.45\textwidth]{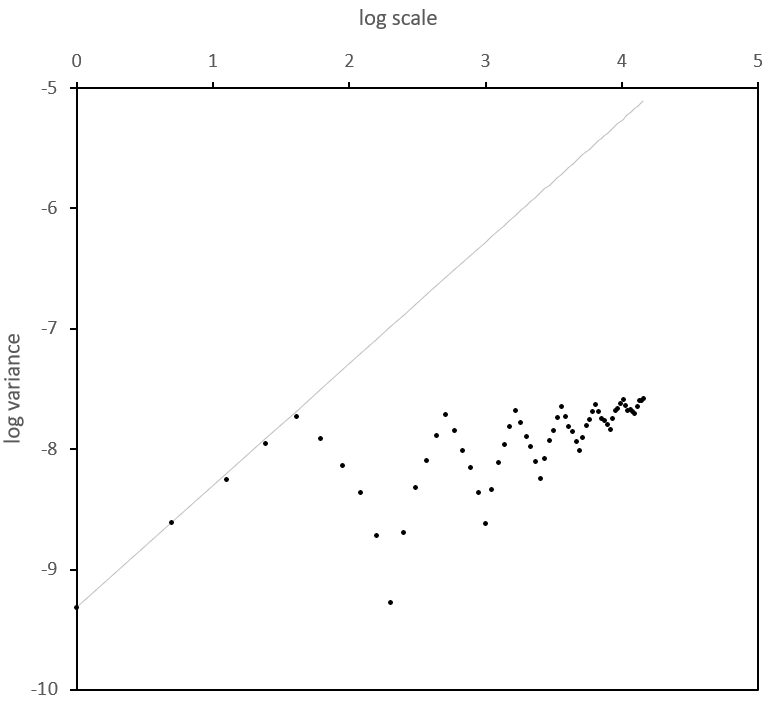}
\begin{minipage}{0.7\textwidth}\caption{Estimated entropy $H^{L+1}_m$ (top left), estimated market information $I^{L+1}_m$ (top right), and estimated partial market information $\mathcal I^{L+1}_m$ (bottom left) for $L\in\llbracket 1,7\rrbracket$ and $m$ equal to 1 (black), 2 (dark grey), and 3 (light grey). The thick dotted lines are derived from the $95\%$ confidence bound of a zero market information for $m=1$. Bottom right: Log-log plot (black) and linear model fitted on the smallest scales (grey). The price series is the simulated pseudo-periodic process of length 3,000 introduced in equation~\eqref{eq:toymodel}.}
	\label{fig:Period}
\end{minipage}
\end{figure}

Figure~\ref{fig:Period} displays the same quantities for a toy model consisting in a pseudo-periodic time series, where price returns follow
\begin{equation}\label{eq:toymodel}
R_{1,i}=\beta R_{1,i-\tau}+\sqrt{1-\beta^2}\varepsilon_i,
\end{equation}
where the $\varepsilon_i$ are i.i.d. standard Gaussian variables, and where we consider $\tau=5$ and $\beta=-0.9$. For this model, the mapping $L\mapsto H^L_m$ shows a clear concavity. The information is mainly provided by the information lagged by 5 units of time, as one can see in the partial information graph. When increasing the time unit $m$, the peak in the partial market information appears for a lower number of lags $L$. Last, the log-log plot is not linear, because of the pseudo-period, illustrating the complementarity of the fractal and informational methods.

\subsection{Empirical analysis}

We now apply the multiscale analysis of market information to real financial data. We focus on four datasets corresponding to various asset classes: a cryptocurrency, the Bitcoin (BTC-USD), between January 2015 and January 2023, two stock indices, the French CAC 40 index and the Japanese Nikkei 225 index, respectively between January 2012 and January 2023 and between January 2022 and June 2022, and an FX rate, the EUR/USD, between December 2006 and January 2016. Moreover, we consider various time samplings: daily observations for the BTC-USD and the CAC 40 index, intraday prices with a 10-minute sampling for Nikkei 225 index and a 1-minute sampling for EUR/USD.

\begin{figure}[tbp]
	\centering
		\includegraphics[width=0.45\textwidth]{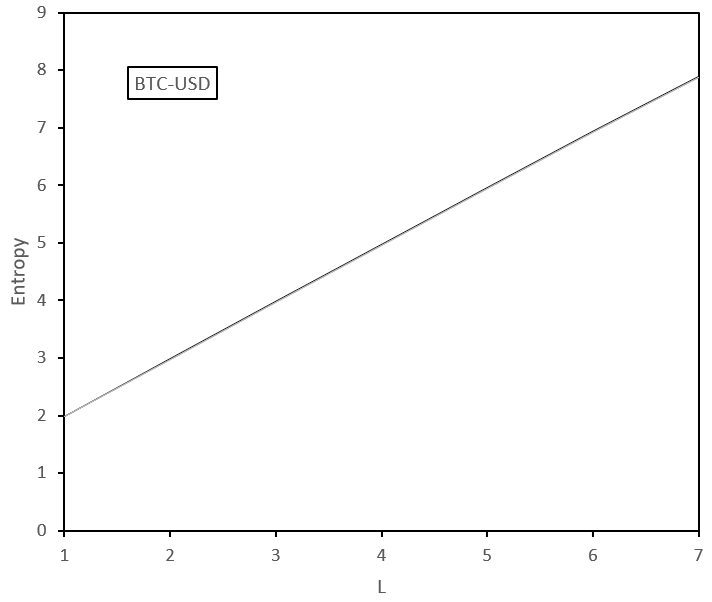} 
		\includegraphics[width=0.45\textwidth]{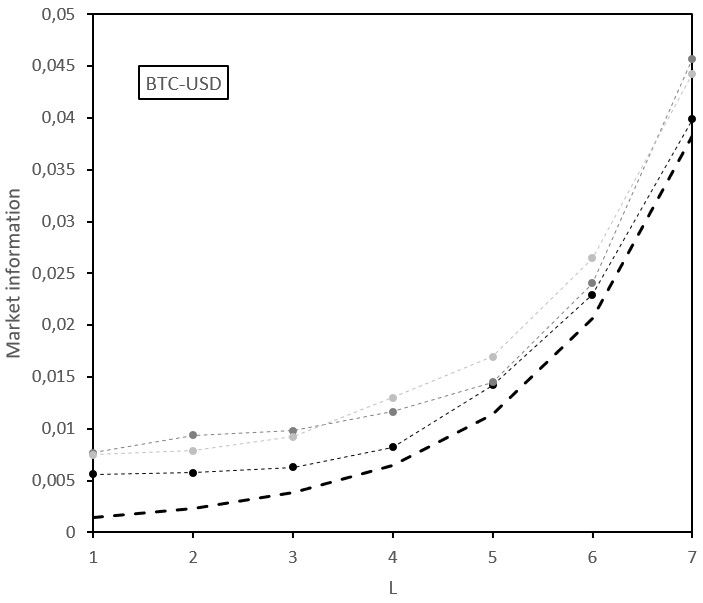} \\
		\includegraphics[width=0.45\textwidth]{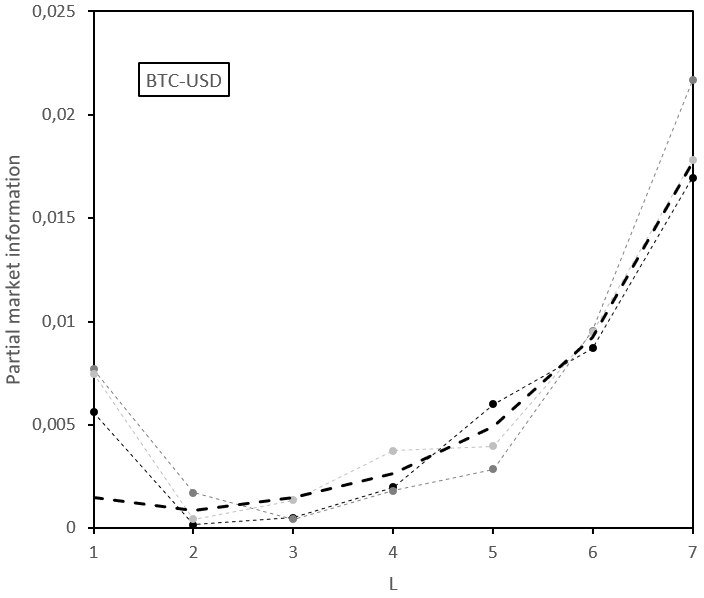}
		\includegraphics[width=0.45\textwidth]{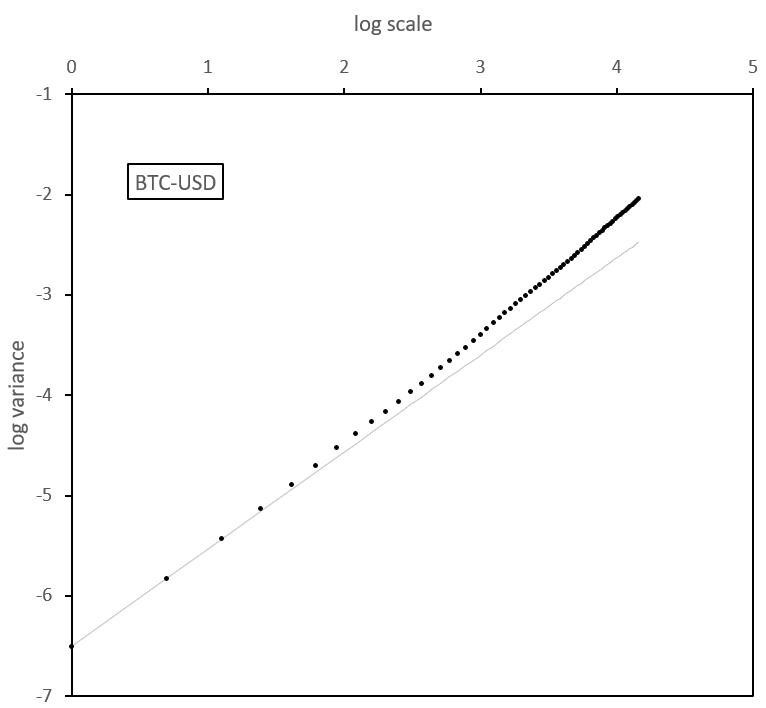}
\begin{minipage}{0.7\textwidth}\caption{Estimated entropy $H^{L+1}_m$ (top left), estimated market information $I^{L+1}_m$ (top right), and estimated partial market information $\mathcal I^{L+1}_m$ (bottom left) for $L\in\llbracket 1,7\rrbracket$ and $m$ equal to 1 (black), 2 (dark grey), and 3 (light grey). The thick dotted lines are derived from the $95\%$ confidence bound of a zero market information for $m=1$. Bottom right: Log-log plot (black) and linear model fitted on the smallest scales (grey). The price series is the BTC-USD, observed at a daily frequency between January 2015 and January 2023, thus representing 2,934 observations.}
	\label{fig:BTC}
\end{minipage}
\end{figure}

We start with the study of the cryptocurrency market, namely the BTC-USD. According to several contributions, the market efficiency of this asset fluctuates through time. In particular, it became progressively less inefficient before 2018 and was inefficient again in 2020~\cite{DGM,KV19,BG,GarcinComplexity}. Since the multiscale approach requires quite long datasets, we do not analyse each year separately but we use this tool to describe the informativeness of the series globally in the period 2015-2023. The Shannon entropy, the market information, the partial market information, and the log-log plot are gathered in Figure~\ref{fig:BTC}. They illustrate that mainly one lag, $L=1$, contributes to the market information, whatever the horizon $m$ considered. Like for the simulations in Section~\ref{sec:Simul}, the estimated market information strongly increases for large values of $L$. But it is the result of the statistical error and not of a true informativeness at these scales. The really relevant quantity to be considered is instead the difference between the market information and the confidence bound of the zero information introduced in Section~\ref{sec:Simul}. Figure~\ref{fig:BTC} thus illustrates the short-term memory of the series of price increase indicators. Last, the log-log plot is not linear but convex, thus confirming the multifractal feature of the time series of Bitcoin prices, documented in other studies~\cite{DKW}. Precisely, a Hurst exponent of 0.48 seems to lead the dynamic at small time scales, whereas a Hurst exponent with a slightly higher value describes the dynamic at greater scales. 

\begin{figure}[tbp]
	\centering
		\includegraphics[width=0.45\textwidth]{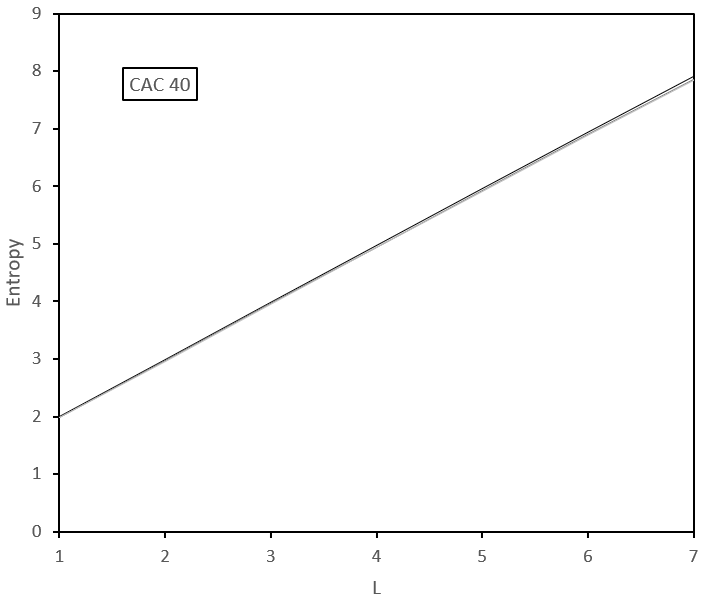} 
		\includegraphics[width=0.45\textwidth]{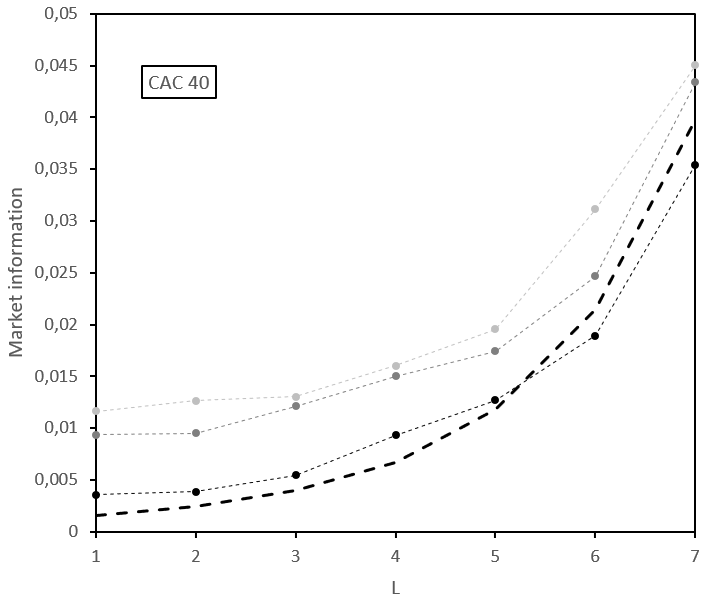} \\
		\includegraphics[width=0.45\textwidth]{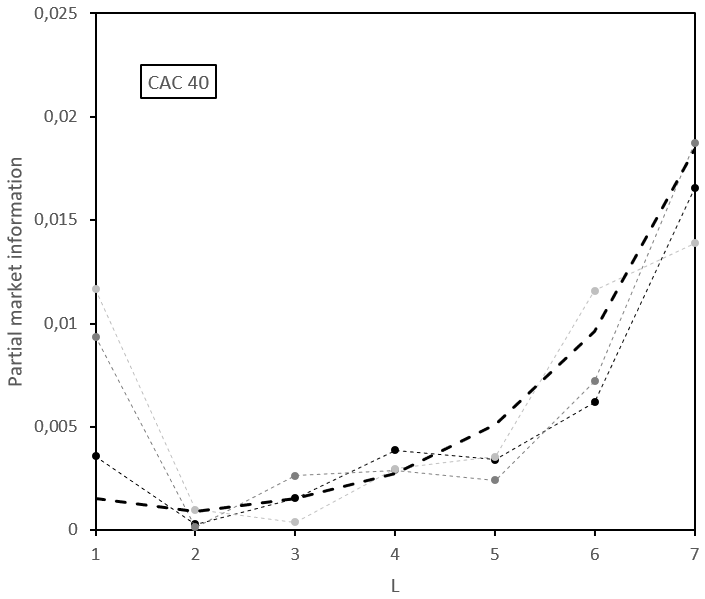}
		\includegraphics[width=0.45\textwidth]{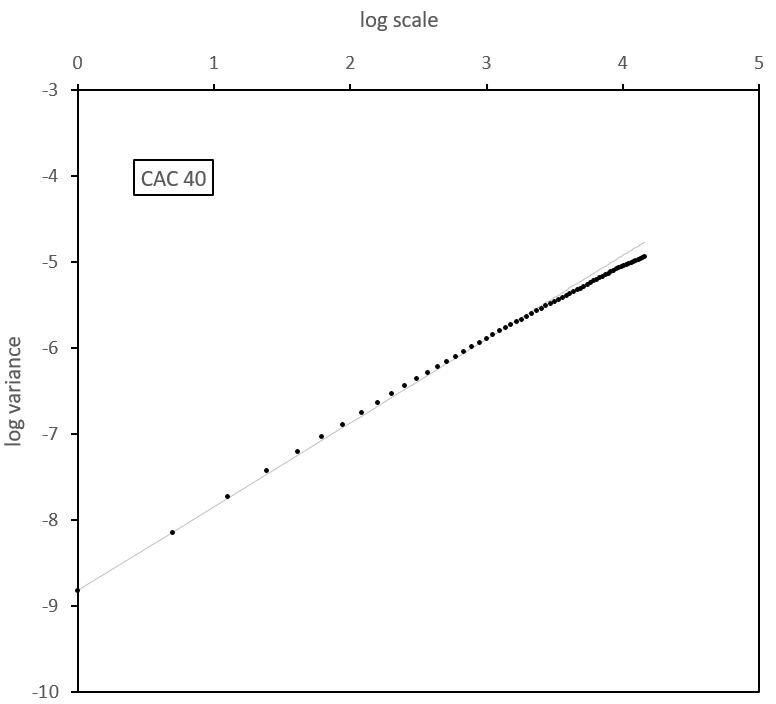}
\begin{minipage}{0.7\textwidth}\caption{Estimated entropy $H^{L+1}_m$ (top left), estimated market information $I^{L+1}_m$ (top right), and estimated partial market information $\mathcal I^{L+1}_m$ (bottom left) for $L\in\llbracket 1,7\rrbracket$ and $m$ equal to 1 (black), 2 (dark grey), and 3 (light grey). The thick dotted lines are derived from the $95\%$ confidence bound of a zero market information for $m=1$. Bottom right: Log-log plot (black) and linear model fitted on the smallest scales (grey). The price series is the CAC 40 index, observed at a daily frequency between January 2012 and January 2023, thus representing 2,822 observations.}
	\label{fig:CAC}
\end{minipage}
\end{figure}

The variations of market efficiency through time for stock indices are also well documented~\cite{Risso,GMF,CLG,ARR,AG,BG}. Focusing on the CAC 40 index, the first lag ($L=1$) seems to be the only significant one, in the information perspective, as one can see in Figure~\ref{fig:CAC}. But, much more markedly than for the BTC-USD, the information content seems greater for longer horizons ($m\in\{2,3\}$) than for $m=1$. It means that the presence of noise makes it more difficult to forecast future price increments at short horizons. This would be consistent with the hypothesis of a trend leading the long-term dynamic of prices. The fractal analysis enriches the description of the price dynamic. Indeed, the log-log plot is concave, depicting a Hurst exponent of 0.49 for the short-term dynamic and a kind of mean-reversion mechanism for very long time scales. 

\begin{figure}[tbp]
	\centering
		\includegraphics[width=0.45\textwidth]{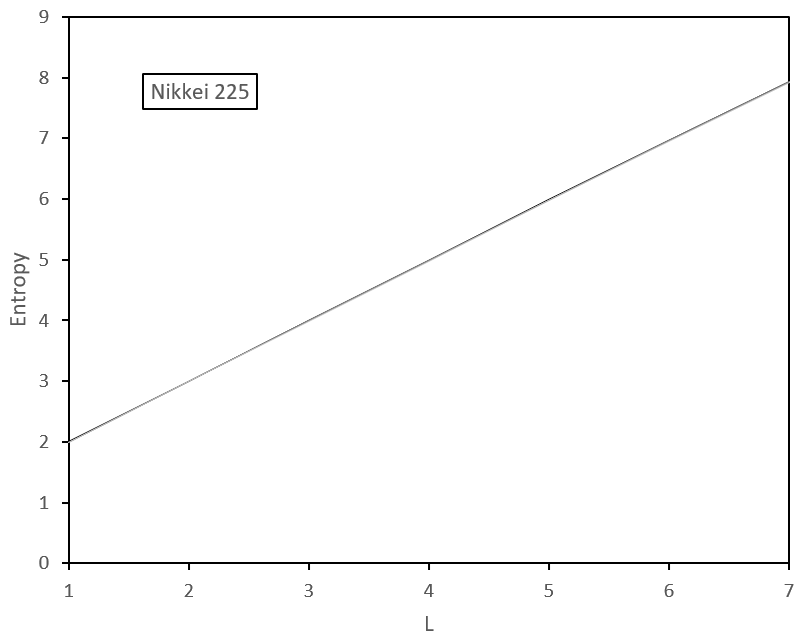} 
		\includegraphics[width=0.45\textwidth]{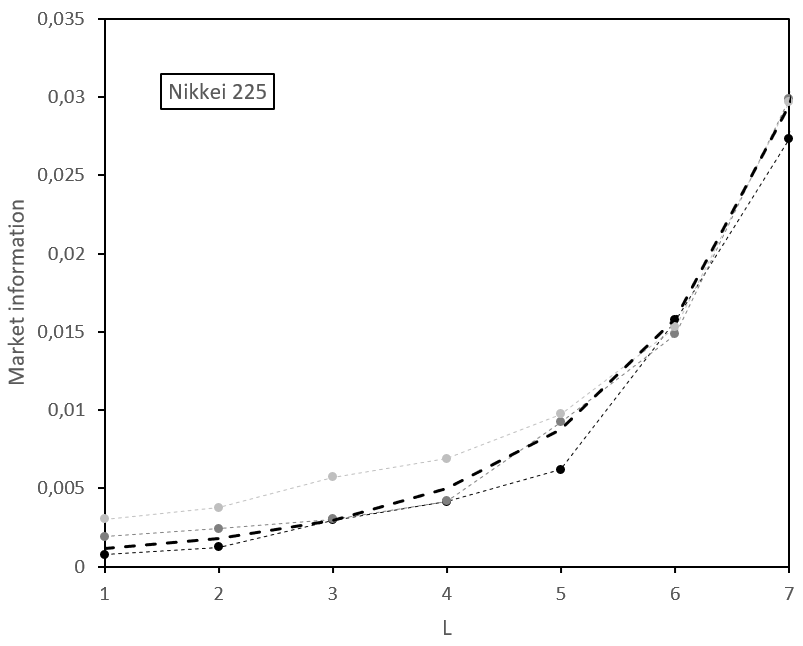} \\
		\includegraphics[width=0.45\textwidth]{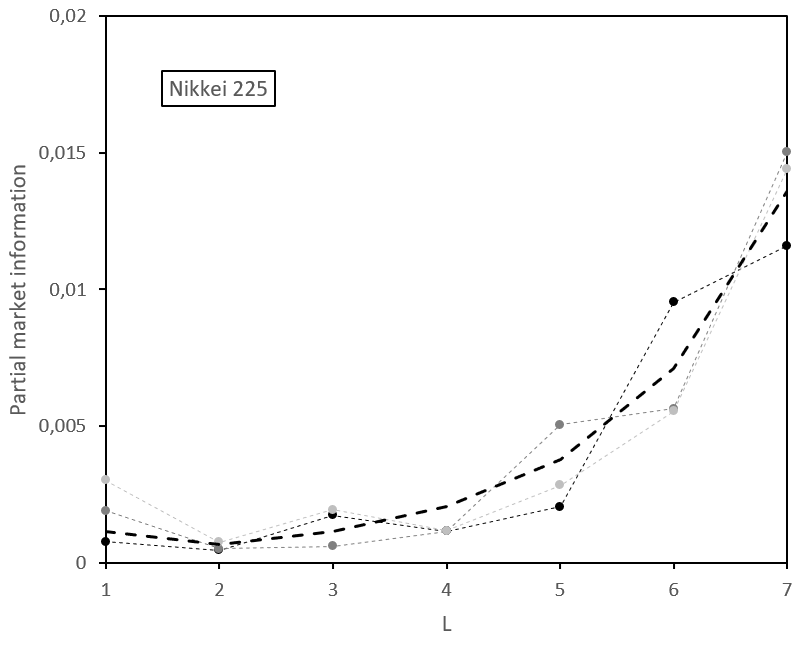}
		\includegraphics[width=0.45\textwidth]{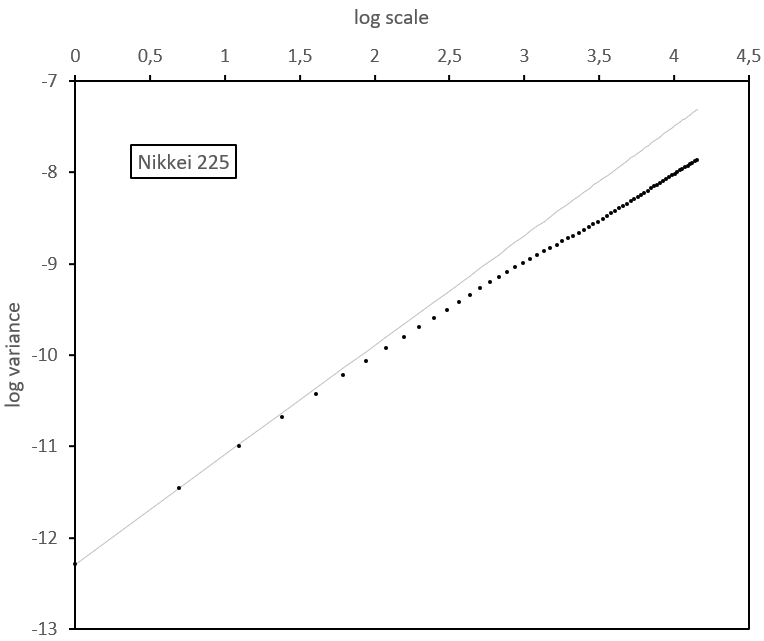}
\begin{minipage}{0.7\textwidth}\caption{Estimated entropy $H^{L+1}_m$ (top left), estimated market information $I^{L+1}_m$ (top right), and estimated partial market information $\mathcal I^{L+1}_m$ (bottom left) for $L\in\llbracket 1,7\rrbracket$ and $m$ equal to 1 (black), 2 (dark grey), and 3 (light grey). The thick dotted lines are derived from the $95\%$ confidence bound of a zero market information for $m=1$. Bottom right: Log-log plot (black) and linear model fitted on the smallest scales (grey). The price series is the close for Nikkei 225, observed each 10 minutes between January 2022 and June 2022, thus representing 3,808 observations.}
	\label{fig:Nikkei}
\end{minipage}
\end{figure}

It may also be interesting to apply the multiscale information analysis to intraday data. Using a 10-minute sampling, the Nikkei 225 index seems to behave quite efficiently at all scales at least for short horizon, that is for $m=1$, according to Figure~\ref{fig:Nikkei}. We get a slightly more significant inefficiency for $m=3$, for which microstructure noise is less noticeable, and for a small number of lagged observations. Like for the CAC 40 index with daily observations, we have a concave log-log plot, but with a higher slope. Indeed, the Hurst exponent for the short-term dynamic is 0.60, indicating forecasting opportunities for time horizons less than about 50 minutes.

\begin{figure}[tbp]
	\centering
		\includegraphics[width=0.45\textwidth]{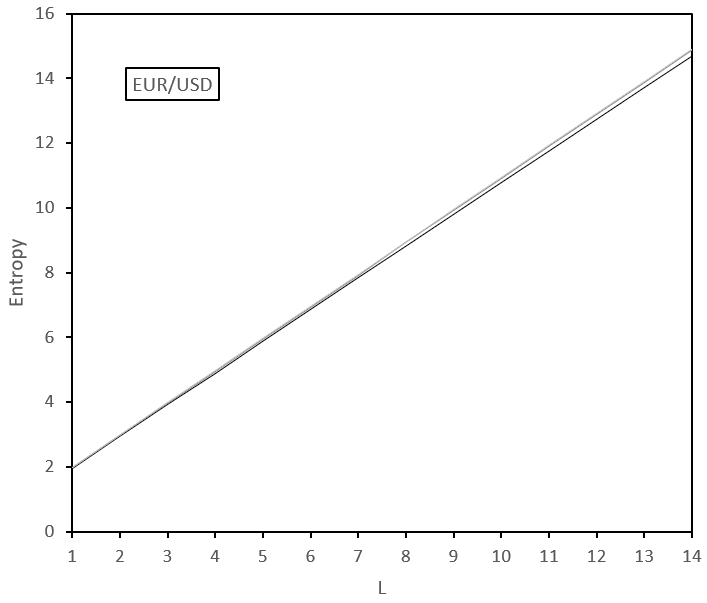} 
		\includegraphics[width=0.45\textwidth]{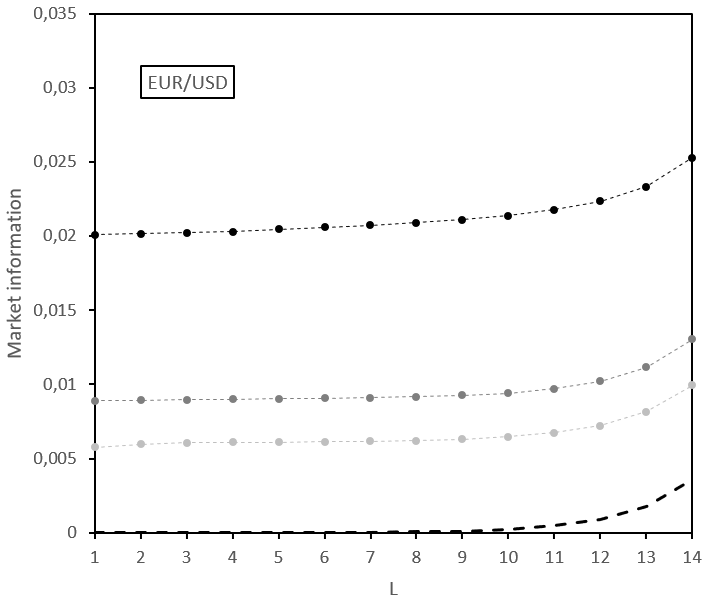} \\
		\includegraphics[width=0.45\textwidth]{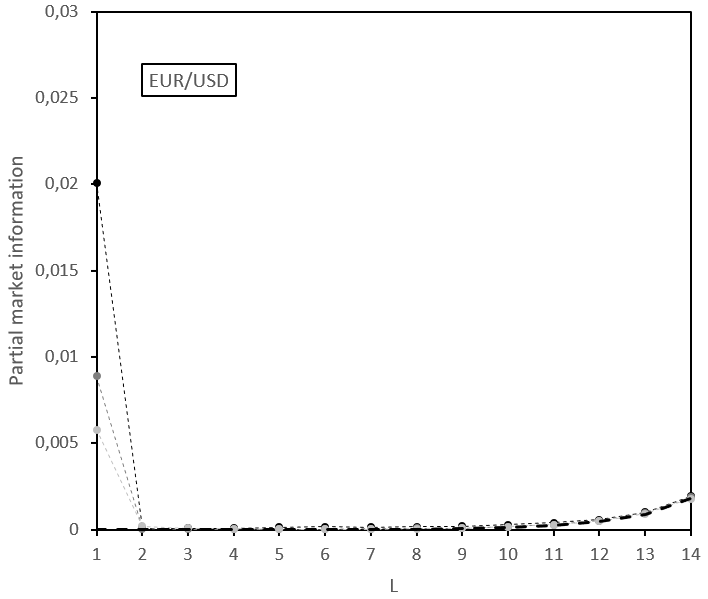}
		\includegraphics[width=0.45\textwidth]{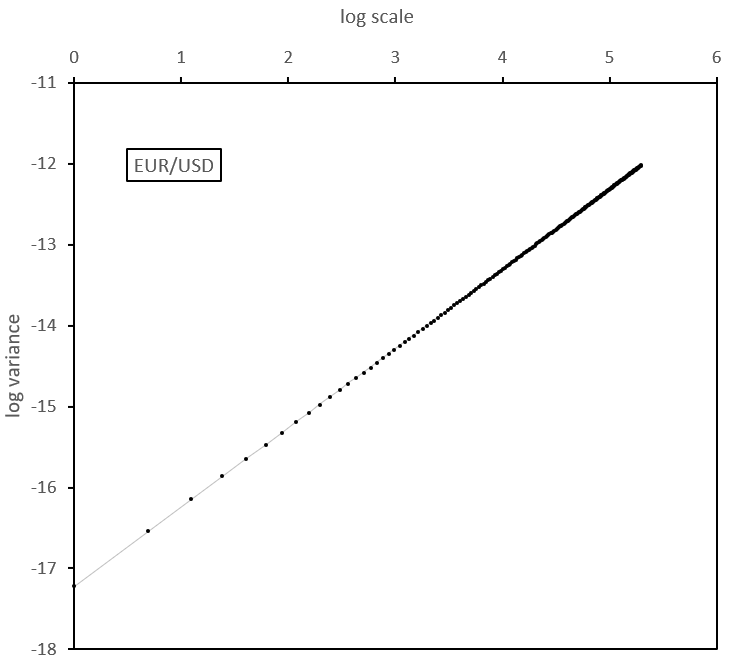}
\begin{minipage}{0.7\textwidth}\caption{Estimated entropy $H^{L+1}_m$ (top left), estimated market information $I^{L+1}_m$ (top right), and estimated partial market information $\mathcal I^{L+1}_m$ (bottom left) for $L\in\llbracket 1,14\rrbracket$ and $m$ equal to 1 (black), 2 (dark grey), and 3 (light grey). The thick dotted lines are derived from the $95\%$ confidence bound of a zero market information for $m=1$. Bottom right: Log-log plot (black) and linear model fitted on the smallest scales (grey). The price series is the close for EUR/USD, observed at each minute between December 2006 and January 2016, thus representing 3,352,284 observations.}
	\label{fig:EURUSD_close}
\end{minipage}
\end{figure}

Each of the real datasets above contains less than 4,000 observations. It limits the analysis to small values of $L$. Using a much longer dataset for the EUR/USD FX rate, sampled every minute, we are able to increase the number of lags considered. The results are gathered in Figure~\ref{fig:EURUSD_close} and indicate the presence of market inefficiency at all the time scales and all the numbers of lags considered, even though the most recent lagged observations has a predominant informativeness. The most significant market information is obtained for $m=1$, with a clear difference with other time scales. Regarding the log-log plot, it is linear, with an estimated Hurst exponent of 0.49, very close to the case of the uninformative standard Brownian motion. It thus seems to contradict the graphs on market information. However, a more deepened study shows that the true dynamic is not Gaussian, so that the standard scaling is the consequence of autocorrelation and fat tails, which compensate each other in the fractal statistic, namely the estimated Hurst exponent~\cite{GarcinMPRE}.

\begin{figure}[tbp]
	\centering
		\includegraphics[width=0.45\textwidth]{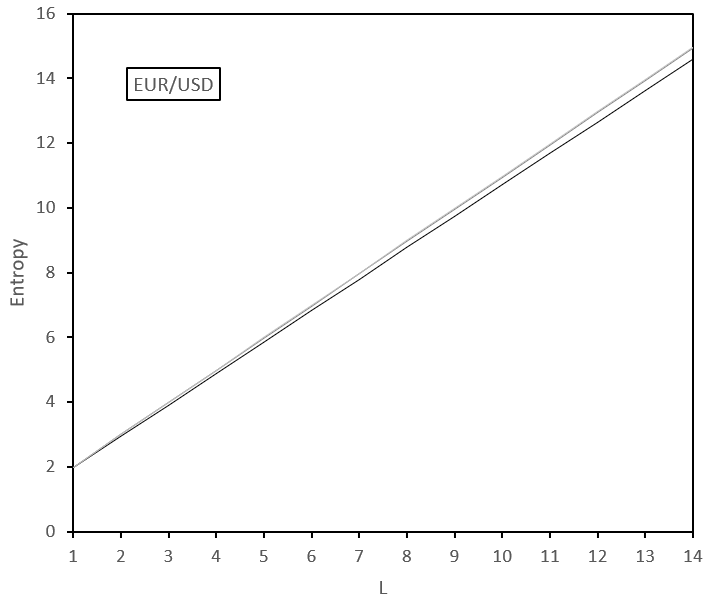} 
		\includegraphics[width=0.45\textwidth]{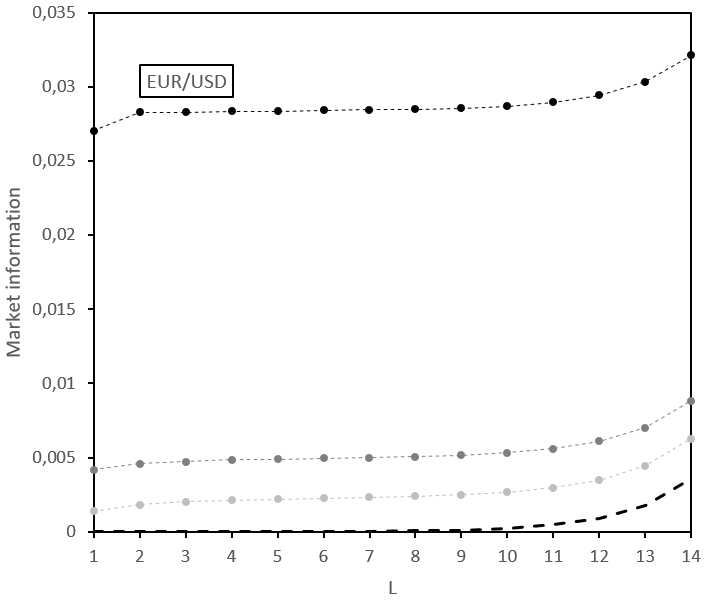} \\
		\includegraphics[width=0.45\textwidth]{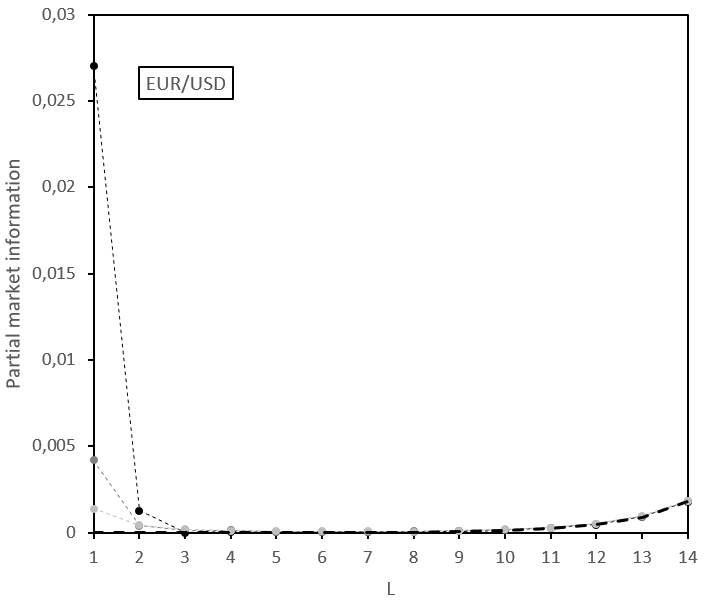}
		\includegraphics[width=0.45\textwidth]{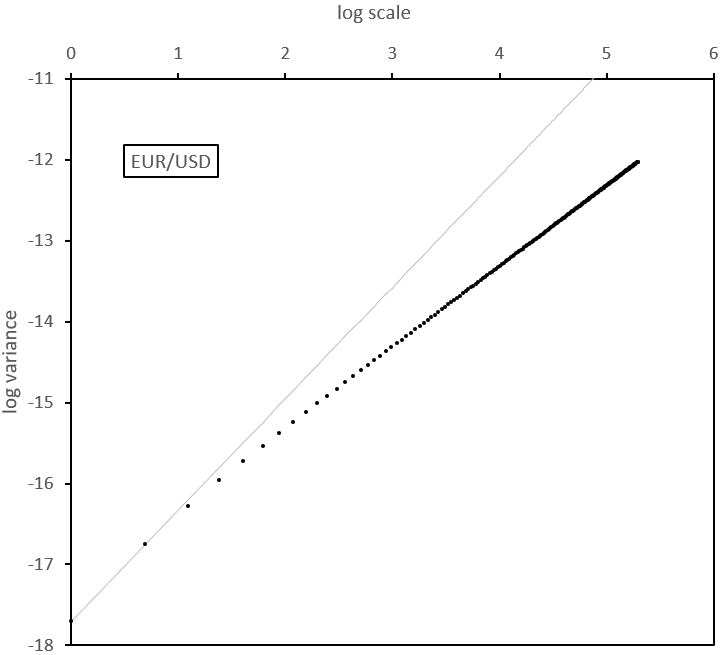}
\begin{minipage}{0.7\textwidth}\caption{Estimated entropy $H^{L+1}_m$ (top left), estimated market information $I^{L+1}_m$ (top right), and estimated partial market information $\mathcal I^{L+1}_m$ (bottom left) for $L\in\llbracket 1,14\rrbracket$ and $m$ equal to 1 (black), 2 (dark grey), and 3 (light grey). The thick dotted lines are derived from the $95\%$ confidence bound of a zero market information for $m=1$. Bottom right: Log-log plot (black) and linear model fitted on the smallest scales (grey). The price series is the average between high and low for EUR/USD, observed in each one-minute interval between December 2006 and January 2016, thus representing 3,352,284 observations.}
	\label{fig:EURUSD_highlow}
\end{minipage}
\end{figure}

All the above results are based on series of close prices. However, it may be difficult to build a trading strategy based on such an analysis. Indeed, the close prices correspond to transaction prices and do not take into account relevant liquidity constraints that a trader would face. If a trading strategy based on the information analysis leads the trader to issue a bid order, it cannot be at the close price if this one is an ask price. In order to get less overoptimistic results, we propose in Figure~\ref{fig:EURUSD_highlow} the same analysis as the one conducted in Figure~\ref{fig:EURUSD_close}, in which we replace the close price of each 1-minute interval by the average between the high and low prices in the same interval. The resulting price is not tradable, but it tends to erase microstructure noise. Surprisingly, the results indicate an even stronger informativeness for $m=1$, whereas the market information slightly decreases for $m\in\{2,3\}$ with respect to the close price case. The log-log plot also shows a stronger slope for small time scales, with the corresponding Hurst exponent estimated at the high value of 0.69.

\section{Conclusion}\label{sec:conclu}

Fractal statistics and information theory provide complementary analysis tools about market efficiency. These two approaches are not unrelated and, in the case where log-prices follow an fBm, we have provided a simple theoretical expression of Shannon's entropy and market information as functions of the Hurst exponent. Since reality may be more complex than the fBm, we have also provided an expression for the market information in the case of another fractal model, namely the delampertized fBm. Combining the information analysis at various time scales and for various sizes of the information set is also a fruitful exercise. Several real-life examples, using daily or intraday price series, have shown the usefulness of this method for understanding the complexity of the price dynamics. Depending on the asset class, one can for instance observe a stronger market information for small time scales or for longer ones. This has some consequences for market practitioners such as market makers, quantitative traders, and risk managers.

\bibliographystyle{plain}

\bibliography{biblioEff}

\appendix

\section{Proof of Theorem~\ref{th:infofBm}}\label{sec:proof_infofBm}

\begin{proof}
Using the notations of Section~\ref{sec:marketInfo}, we write $J_{m,t}=\indic_{\{P_t-P_{t-m}> 0\}}=\indic_{\{B^{\mathcal H}_t-B^{\mathcal H}_{t-m}> 0\}}$. Let $Y=B^{\mathcal H}_{2m}-B^{\mathcal H}_{m}$ and $Z=B^{\mathcal H}_{m}-B^{\mathcal H}_0$. After equations~\eqref{eq:marketInfo} and~\eqref{eq:conditionalEntropy} and the property of stationarity of the increments of an fBm, we have
$$\begin{array}{ccl}
I^2_m & = & 1-H(J_{m,.+m}|J_{m,.}) \\
 & = & 1+\proba(J_{m,.}=1)\left[f\left(\proba(J_{m,.+m}=1|J_{m,.}=1)\right)+f\left(\proba(J_{m,.+m}=0|J_{m,.}=1)\right)\right] \\
 & & +\proba(J_{m,.}=0)\left[f\left(\proba(J_{m,.+m}=1|J_{m,.}=0)\right)+f\left(\proba(J_{m,.+m}=0|J_{m,.}=0)\right)\right] \\
 & = & 1+\proba(Z>0)\left[f\left(\proba(Y>0|Z>0)\right)+f\left(\proba(Y\leq 0|Z>0)\right)\right] \\
 & & +\proba(Z\leq 0)\left[f\left(\proba(Y>0|Z\leq 0)\right)+f\left(\proba(Y\leq 0|Z\leq 0)\right)\right]. 
\end{array}$$
Noting on the one hand that $\proba(Z\leq 0)=\proba(Z> 0)=1/2$ and on the other hand that the events $Y>0$ and $Y\leq 0$ are complementary, we get
\begin{equation}\label{eq:proofinfofBm}
\begin{array}{ccl}
I^2_m & = & 1+\frac{1}{2}\left[f\left(\proba(Y>0|Z>0)\right)+f\left(1-\proba(Y>0|Z>0)\right)\right] \\
 & & +\frac{1}{2}\left[f\left(\proba(Y>0|Z\leq 0)\right)+f\left(1-\proba(Y>0|Z\leq 0)\right)\right].
\end{array}
\end{equation}
The vector $(Y,Z)$ is Gaussian of mean zero and of covariance $\Sigma_{YZ}$, such that:
$$\Sigma_{YZ}=m^{2\mathcal H}\sigma^2\left(\begin{array}{cc}
1 & \rho \\
\rho & 1
\end{array}\right),$$
where $\rho=2^{2\mathcal H-1}-1$. Indeed, exploiting equation~\eqref{eq:covFBM}, we get:
$$\begin{array}{ccl}
\Cov(Y,Z) & = & \E(B^{\mathcal H}_{2m}B^{\mathcal H}_m)-\E(B^{\mathcal H}_{2m}B^{\mathcal H}_0)-\E((B^{\mathcal H}_m)^2)+\E(B^{\mathcal H}_mB^{\mathcal H}_0) \\
 & = & \frac{\sigma^2}{2} \left((2m)^{2\mathcal H}-2m^{2\mathcal H}\right)=m^{2\mathcal H}\sigma^2\rho.
\end{array}$$
Simple matrix manipulations provide the determinant $|\Sigma_{YZ}|=m^{4\mathcal H}\sigma^4(1-\rho^2)$ and 
$$\Sigma_{YZ}^{-1}=\frac{1}{m^{2\mathcal H}\sigma^2(1-\rho^2)}\left(\begin{array}{cc}
1 & -\rho \\
-\rho & 1
\end{array}\right).$$
As a consequence, we can calculate the following joint probability, thanks to the successive substitutions $u=(y-\rho z)/m^{\mathcal H}\sigma\sqrt{1-\rho^2}$ and $v=z/m^{\mathcal H}\sigma$:
$$\begin{array}{ccl}
\proba(Y> 0,Z\leq 0) & = & \int_{-\infty}^0\int_{0}^{+\infty} \frac{1}{2\pi |\Sigma_{YZ}|^{1/2}} \exp\left(-\frac{[y\ z]\Sigma_{YZ}^{-1}[y\ z]^T}{2}\right) dy dz \\
 & = & \frac{1}{2\pi m^{2\mathcal H}\sigma^2\sqrt{1-\rho^2}} \int_{-\infty}^0\int_{0}^{+\infty} \exp\left(-\frac{(y-\rho z)^2+(1-\rho^2)z^2}{2m^{2\mathcal H}\sigma^2(1-\rho^2)}\right) dy dz \\
 & = & \frac{1}{2\pi m^{\mathcal H}\sigma} \int_{-\infty}^0\int_{-\rho z/m^{\mathcal H}\sigma\sqrt{1-\rho^2}}^{+\infty} \exp\left(-\frac{u^2}{2}\right) \exp\left(-\frac{z^2}{2m^{2\mathcal H}\sigma^2}\right) du dz \\
 & = & \frac{1}{2\pi} \int_{-\infty}^0 \left(1-N\left(\frac{-\rho v}{\sqrt{1-\rho^2}}\right)\right) \exp\left(-\frac{v^2}{2}\right) du dv \\
 & = & \frac{1}{2\pi} \int_0^{\infty} N\left(\frac{-\rho v}{\sqrt{1-\rho^2}}\right) \exp\left(-\frac{v^2}{2}\right) dv.
\end{array}$$
This is equal to $\frac{1}{4}-\frac{1}{2\pi}\arctan\left(\frac{\rho}{\sqrt{1-\rho^2}}\right)$~\cite[Lemma 1]{GarcinForecast}. Noting that $\proba(Z\leq 0)=1/2$, we finally get:
$$\proba(Y> 0|Z\leq 0)=\frac{1}{2}-\frac{1}{\pi}\arctan\left(\frac{\rho}{\sqrt{1-\rho^2}}\right).$$
Similarly
$$\proba(Y> 0|Z> 0)=\frac{1}{2}+\frac{1}{\pi}\arctan\left(\frac{\rho}{\sqrt{1-\rho^2}}\right).$$
Going back to equation~\eqref{eq:proofinfofBm}, we thus get
$$I_m^2=1+f\left(\frac{1}{2}-\frac{1}{\pi}\arctan\left(\frac{\rho}{\sqrt{1-\rho^2}}\right)\right)+f\left(\frac{1}{2}+\frac{1}{\pi}\arctan\left(\frac{\rho}{\sqrt{1-\rho^2}}\right)\right).$$
\end{proof}

\section{Proof of Theorem~\ref{th:infoLamperti}}\label{sec:proof_infoLamperti}

\begin{proof}
The process $\mathcal X^{\mathcal H,\theta}_t$ is Gaussian with stationary increments~\cite{GarcinLamperti}. Noting $Y=\mathcal X^{\mathcal H,\theta}_{2m}-\mathcal X^{\mathcal H,\theta}_m$ and $Z=\mathcal X^{\mathcal H,\theta}_m-\mathcal X^{\mathcal H,\theta}_0$, we have~\cite{GarcinLamperti} 
$$\Var(Y)=\Var(Z)=\sigma^2\left(2-h(m\theta)\right)$$
and
$$\Cov(Y,Z)=\frac{\sigma^2}{2}\left(-h(2m\theta)+2h(m\theta)-2\right),$$
so that
$$\corr(Y,Z)=-1+\frac{2-h(2m\theta)}{4-2h(m\theta)}.$$
One sees immediately in the proof of Theorem~\ref{th:infofBm} that equation~\eqref{eq:infofBm}, written for the sole fBm, is true for any Gaussian process with stationary increments as soon as one replaces the parameter $\rho$ by the correlation between the increments $Y$ and $Z$. This leads to Theorem~\ref{th:infoLamperti}.
\end{proof}

\section{Proof of Theorem~\ref{th:concaventropy}}\label{sec:proof_concaventropy}

\begin{proof}
We know by equation~\eqref{eq:chainrule} that  
$$H^{L+1}_m=H^{L}_m+H(J_{m,.+Lm}|J_{m,.},...,J_{m,.+(L-1)m}).$$
Moreover, the chain rule also leads to~\cite[Th.2.2.1 and Th.2.5.1]{CT}
$$H^{L+2}_m=H^{L}_m+H(J_{m,.+(L+1)m}|J_{m,.},...,J_{m,.+Lm})+H(J_{m,.+Lm}|J_{m,.},...,J_{m,.+(L-1)m}).$$ 
Therefore, 
$$H^{L+2}_m-H^{L+1}_m=H(J_{m,.+(L+1)m}|J_{m,.},...,J_{m,.+Lm})\geq 0.$$ 
So the considered mapping is increasing. Then, the second difference of the function $L\mapsto H^{L}_m$ is
$$\begin{array}{ccl}
H^{L+2}_m-2H^{L+1}_m+H^L_m & = & H(J_{m,.+(L+1)m}|J_{m,.},...,J_{m,.+Lm}) - H(J_{m,.+Lm}|J_{m,.},...,J_{m,.+(L-1)m}) \\
 & = & H(J_{m,.+(L+1)m}|J_{m,.},...,J_{m,.+Lm}) - H(J_{m,.+(L+1)m}|J_{m,.+m},...,J_{m,.+Lm}) \\
 & \leq & 0
\end{array}$$ 
where we justify the penultimate line by the stationarity of the series and the last one with the argument that conditioning reduces entropy~\cite[Th.2.6.5]{CT}. The second difference of $L\mapsto H^{L}_m$ being negative, it is a concave function.
\end{proof}

\end{document}